\documentclass[aps,prb,reprint,superscriptaddress,longbibliography,floatfix]{revtex4-2}

\usepackage{graphicx}
\usepackage{dcolumn}
\usepackage{bm}
\usepackage{amsmath,amssymb}

\begin{document}

\title{Continuum model for the terahertz dielectric response of glasses}

\author{Tatsuya Mori}
\email{mori@ims.tsukuba.ac.jp}
\affiliation{Department of Materials Science, University of Tsukuba, 1-1-1 Tennodai, Tsukuba, Ibaraki 305-8573, Japan}

\author{Hideyuki Mizuno}
\affiliation{Graduate School of Arts and Sciences, The University of Tokyo, 3-8-1 Komaba, Meguro-ku, Tokyo 153-8902, Japan}

\author{Yuzuki Motokawa}
\affiliation{Department of Materials Science, University of Tsukuba, 1-1-1 Tennodai, Tsukuba, Ibaraki 305-8573, Japan}

\author{Dan Kyotani}
\affiliation{Department of Materials Science, University of Tsukuba, 1-1-1 Tennodai, Tsukuba, Ibaraki 305-8573, Japan}

\author{Soo Han Oh}
\affiliation{Department of Materials Science, University of Tsukuba, 1-1-1 Tennodai, Tsukuba, Ibaraki 305-8573, Japan}

\author{Yasuhiro Fujii}
\affiliation{Institute for Open and Transdisciplinary Research Initiatives, Osaka University, 2-1 Yamada-oka, Suita, Osaka 565-0871, Japan}
\affiliation{Research Organization of Science and Technology, Ritsumeikan University, 1-1-1 Noji-higashi, Kusatsu, Shiga 525-8577, Japan}

\author{Akitoshi Koreeda}
\affiliation{Department of Physical Sciences, Ritsumeikan University, 1-1-1 Noji-higashi, Kusatsu, Shiga 525-8577, Japan}

\author{Shinji Kohara}
\affiliation{Center for Basic Research on Materials, National Institute for Materials Science (NIMS), 1-2-1 Sengen, Tsukuba, Ibaraki 305-0047, Japan}

\author{Seiji Kojima}
\affiliation{Department of Materials Science, University of Tsukuba, 1-1-1 Tennodai, Tsukuba, Ibaraki 305-8573, Japan}

\begin{abstract}
Boson peak dynamics in glasses produce a robust crossover in the terahertz (THz) dielectric response that standard Debye or Lorentz models do not capture. We develop a continuum description of this THz response, coupling an infrared-effective charge fluctuation spectrum to a frequency-dependent shear modulus, and apply it to glycerol glass. The model reproduces the measured complex dielectric function and the nearly linear infrared light-vibration coupling around the boson peak, and highlights the dominant role of transverse shear dynamics.
\end{abstract}

\maketitle

\section{Introduction}
In disordered solids, a universal excess of vibrational modes appears in the terahertz (THz) range---the
boson peak (BP)---which is observed as a peak in the reduced vibrational density of states
(VDOS) $g(\omega)/\omega^{D-1}$ (where $D$ is the spatial dimension) relative to the Debye level~\cite{Nakayama2002,Ramos2022}.
Despite decades of work, the microscopic origin of the BP remains an open problem that
continues to stimulate experiments~\cite{Zeller1971,Buchenau1986,Ruffle2008,Baldi2010,Chumakov2011,Phillips1981,Malinovsky1986,Kojima1993,Kabeya2016,Mori2020},
simulations~\cite{Shintani2008,Tanguy2010,Franz2015,Lerner2016,Mizuno2017,Hu2022,Lerner2023},
and theory~\cite{Schirmacher2006,Schirmacher2024,Wyart2010,DeGiuli2014,Buchenau1991,Gurevich2003,Rainone2021,Moriel2024}.
The BP is tied to several hallmark anomalies of glasses: unusually low thermal conductivity~\cite{Zeller1971},
nanoscale plasticity~\cite{Tanguy2010}, and a marked decrease in transmittance in the THz region above the BP
frequency~\cite{Kabeya2016,Mori2020}.
Understanding the BP and its coupling to light is essential for fundamental physics and for
predicting and controlling THz optical properties.
In particular, emerging THz communication technologies demand window and substrate materials
with low permittivity and low loss~\cite{Ako2020}.
Motivated by these considerations, we focus on the dielectric response in the THz regime and
develop a framework that connects the observed absorption to the underlying vibrational dynamics.
Empirically, the THz dielectric response $\varepsilon(\omega)$ in glasses shows a robust pattern that
standard models fail to capture: below the BP frequency $\omega_{\rm BP}$ the response is
resonance-like (Lorentz-oscillator-like), whereas above $\omega_{\rm BP}$ it becomes relaxational
(Debye-like)~\cite{Kabeya2016,Mori2020}.
Neither standard relaxational dielectric functions---Debye and common empirical generalizations
(e.g., the Havriliak--Negami form~\cite{Havriliak1967})---nor an overdamped Lorentz oscillator reproduces this crossover;
a unified analytic form for $\varepsilon(\omega)$ that captures this universal crossover has been lacking.

Several beyond-Debye/Lorentz approaches have been pursued for the THz dielectric anomaly
around the BP.
Zaccone and co-workers formulated a phonon-polaritonic response by extending
a Lorentz oscillator with a wavenumber-dependent damping $\Gamma(k)$, but the choice of $\Gamma(k)$ limits transferability across different glasses~\cite{Casella2021}.
A different route links dielectric relaxation to viscoelasticity: in the
Gemant--DiMarzio--Bishop model the constant Debye viscosity is replaced by a
frequency-dependent one, which can be expressed via a complex shear modulus~\cite{DiMarzioBishop1974,Niss2005}.
However, the model was devised for liquid-side relaxation and carries an explicit temperature dependence, making a direct application to low-temperature glass spectra nontrivial.
Along this line, Zeitler \textit{et al.} discussed the scaling
$g(\omega)/\omega^{2} \propto \alpha(\omega)/\omega^{3}$ (with $\alpha(\omega)$ the absorption coefficient)~\cite{Kolbel2023}; nevertheless, a unified,
material-independent $\varepsilon(\omega)$ capturing the BP crossover remains elusive.

Light--vibration coupling coefficients in the BP region are spectroscopic-channel dependent.
Hyper-Raman scattering (HRS) studies have shown an approximately constant $C_{\rm HRS}(\omega)$ around the BP in vitreous silica~\cite{Hehlen2000HRS,HehlenSimon2012JRS} and have highlighted the tensor-rank dependence of the coupling in network glasses~\cite{Simon2006HRS,HehlenSimon2012JRS}, whereas ordinary Raman measurements have reported a nearly linear $C_{\rm Raman}(\omega)$ near the BP in a broad set of glasses~\cite{SurovtsevSokolov2002Raman}; analysis of our Raman data for glycerol likewise yields an approximately linear $C_{\rm Raman}(\omega)$ in the BP region (see Appendix~\ref{app:raman-coupling}).
For the infrared (IR) channel considered here, Taraskin \textit{et al.}, building on Maradudin's
linear-response theory for optical phonons in crystals~\cite{Maradudin1961} and decomposing the effective atomic-charge fluctuations into correlated (crystal-like) and uncorrelated (disorder-induced) parts, derived the analytic form for the IR light--vibration coupling coefficient $C_{\rm IR}(\omega)=\alpha(\omega)/g(\omega)=A+B\omega^2$, which reproduces the IR absorption of silica and several other oxide glasses~\cite{Taraskin2006}.
However, for glycerol the experimentally extracted $C_{\rm IR}(\omega)$ displays a pronounced linear dependence in the vicinity of the BP, clearly deviating from $A+B\omega^2$ (see Fig.~\ref{fig:reduced-spectra-ir-coupling}(b)).
Similar linear trends have been reported for a broad range of amorphous materials, including hydrogen-bonded~\cite{Kabeya2016} and polymeric glasses~\cite{Hashimoto2016}, proteins~\cite{Mori2020}, and even some inorganic glasses~\cite{Taraskin2007}.
These IR observations call for a formulation that explains the linear $C_{\rm IR}(\omega)$ while remaining consistent with the measured complex dielectric function $\varepsilon(\omega)$ (both its real and imaginary parts, $\varepsilon'(\omega)$ and $\varepsilon''(\omega)$).

In this work, we introduce a continuum model that explicitly incorporates the IR-effective
charge fluctuation spectrum $\Delta q(k)$ and the effective complex shear modulus $G(\omega)$.
With a low-$k$ form $\Delta q(k)\approx q_0+q_2 k^2$, the model simultaneously reproduces
$\varepsilon'(\omega)$ and $\varepsilon''(\omega)$ and accounts for the linear frequency dependence
of $C_{\rm IR}(\omega)$ in the BP region, in quantitative agreement with experimental data.
Moreover, by increasing the constant-to-quadratic ratio $R \equiv q_{0}/(q_{2}k_{\mathrm{D}}^{2})$ ($k_{\mathrm{D}}$ is the Debye wavenumber), the response continuously approaches a Taraskin-like form, so the present framework both explains the linear regime and contains the Taraskin expression as a limiting case.

\section{Continuum model}

\subsection{Microscopic origin of the electric driving term}

Before writing the continuum equation of motion, we briefly derive the electric source term from the microscopic rigid-ion dipole.
In a rigid-ion picture, each atom or site $i$ is characterized by a time-independent rigid-ion charge $q_i$, an equilibrium position $\mathbf{R}_{i}^{0}$, and a displacement $\mathbf{u}_{i}(t)$ from $\mathbf{R}_{i}^{0}$.
Its instantaneous position is then
\begin{equation}
\label{eq:main-position}
\mathbf{R}_{i}(t)=\mathbf{R}_{i}^{0}+\mathbf{u}_{i}(t),
\end{equation}
and the microscopic dipole moment is the vector
\begin{equation}
\label{eq:main-dipole-decomposition}
\begin{aligned}
\mathbf{d}(t)
&=\sum_i q_i\mathbf{R}_{i}(t) \\
&=\sum_i q_i\mathbf{R}_{i}^{0}
 +\sum_i q_i\mathbf{u}_{i}(t)
 \equiv \mathbf{d}_{0}+\mathbf{d}^{\rm vib}(t).
\end{aligned}
\end{equation}
Here $\mathbf{d}_{0}$ is the static dipole associated with the equilibrium reference configuration, and $\mathbf{d}^{\rm vib}(t)$ is the vibrational contribution to the dipole.
The first term is static and does not generate a force on the vibrational coordinates; the field-induced vibrational response is governed by
\begin{equation}
\label{eq:main-dvib-discrete}
\mathbf{d}^{\rm vib}(t)=\sum_i q_i\mathbf{u}_{i}(t).
\end{equation}
The corresponding local IR-effective charge weight in the equilibrium reference configuration may be decomposed as
$q(\mathbf{r})=q_{\rm av}+\Delta q(\mathbf{r})$, where $q_{\rm av}$ is its spatially uniform component.
For the charge-neutral rigid-ion system considered here, $q_{\rm av}=0$, leaving the fluctuation field $\Delta q(\mathbf{r})$.
The continuum-field representation of this discrete sum is therefore
\begin{equation}
\label{eq:main-dvib-continuum}
\mathbf{d}^{\rm vib}(t)
\simeq
\int d^{3}r\,\Delta q(\mathbf{r})\,\mathbf{u}(\mathbf{r},t).
\end{equation}
Here $\mathbf{r}$ denotes a position in the equilibrium reference configuration, $\mathbf{u}(\mathbf{r},t)$ is the continuum displacement field, and $\Delta q(\mathbf{r})$ denotes the local IR-effective charge fluctuation field in that reference configuration.
Equation~\eqref{eq:main-dvib-continuum} represents the discrete sum in Eq.~\eqref{eq:main-dvib-discrete} in continuum-field form: the spatial dependence of the IR-effective charge fluctuations is retained in $\Delta q(\mathbf{r})$, while the site displacements $\mathbf{u}_{i}(t)$ are represented by $\mathbf{u}(\mathbf{r},t)$.
The approximate equality in Eq.~\eqref{eq:main-dvib-continuum} reflects this discrete-to-continuum replacement.
The interaction energy between a spatially uniform external electric field $\mathbf{E}(t)$ and this vibrational dipole is
\begin{equation}
\label{eq:main-interaction-energy}
\begin{aligned}
H_{\rm int}^{\rm vib}(t)
&=-\mathbf{E}(t)\cdot\mathbf{d}^{\rm vib}(t) \\
&=-\int d^{3}r\,\Delta q(\mathbf{r})\,
\mathbf{E}(t)\cdot\mathbf{u}(\mathbf{r},t).
\end{aligned}
\end{equation}
Taking the functional derivative with respect to the displacement field $\mathbf{u}(\mathbf{r},t)$ gives the electric force density
\begin{equation}
\label{eq:main-electric-force-density}
\mathbf{f}_{E}(\mathbf{r},t)
=-\frac{\delta H_{\rm int}^{\rm vib}}{\delta\mathbf{u}(\mathbf{r},t)}
=\Delta q(\mathbf{r})\mathbf{E}(t).
\end{equation}
After Fourier transforming in time, this becomes
$\mathbf{f}_{E}(\mathbf{r},\omega)=\Delta q(\mathbf{r})\mathbf{E}(\omega)$,
where $\omega$ denotes angular frequency and $\mathbf{E}(\omega)$ is the corresponding Fourier component of the external electric field.
This is the source term used in Eq.~\eqref{eq:eom}.
Appendix~\ref{app:electric-driving} gives the corresponding discrete-to-continuum derivation of this electric source term.

\subsection{Equation of motion and dielectric response}

We describe the THz dielectric response of a glassy solid using an isotropic viscoelastic continuum with bulk modulus $K$, complex shear modulus $G(\omega)$, and the local IR-effective charge fluctuation field $\Delta q(\mathbf{r})$ introduced above. In the frequency domain, the displacement field $\mathbf{u}(\mathbf{r},\omega)$ driven by a spatially uniform THz electric field $\mathbf{E}(\omega)=E(\omega)\hat{\mathbf{e}}$ obeys
\begin{equation}
\label{eq:eom}
\begin{aligned}
-\rho\omega^{2}\mathbf{u}(\mathbf{r},\omega)
&=\left(K+\frac{4}{3}G(\omega)\right)\nabla\!\left(\nabla\!\cdot\!\mathbf{u}(\mathbf{r},\omega)\right)\\
&\quad -G(\omega)\nabla\times\!\left(\nabla\times\mathbf{u}(\mathbf{r},\omega)\right)\\
&\quad +\Delta q(\mathbf{r})\mathbf{E}(\omega).
\end{aligned}
\end{equation}
Here $\rho$ is the mass density. We use the Fourier representation
\begin{equation}
\label{eq:main-fourier-representation}
\begin{aligned}
\mathbf{u}(\mathbf{r},\omega)
&=\frac{1}{\sqrt{V}}\sum_{\mathbf{k},\alpha}u_{\alpha}(\mathbf{k},\omega)\,
\hat{\mathbf{e}}_{\alpha}(\mathbf{k})e^{\mathrm{i}\mathbf{k}\cdot\mathbf{r}}, \\
\Delta q(\mathbf{r})
&=\frac{1}{\sqrt{V}}\sum_{\mathbf{k}}\Delta q(\mathbf{k})e^{\mathrm{i}\mathbf{k}\cdot\mathbf{r}},
\end{aligned}
\end{equation}
where $\Delta q(\mathbf{r})$ is real, so that $\Delta q(-\mathbf{k})=\Delta q(\mathbf{k})^{*}$. For each wave vector, the three acoustic branches $\alpha=L,T_1,T_2$ have polarization vectors $\hat{\mathbf{e}}_{\alpha}(\mathbf{k})$ ($\hat{\mathbf{e}}_{L}(\mathbf{k})=\mathbf{k}/|\mathbf{k}|$, $\hat{\mathbf{e}}_{T_i}(\mathbf{k})\cdot\mathbf{k}=0$). The uniform field couples to branch $\alpha$ through the scalar projection $E_{\alpha}(\omega)=\hat{\mathbf{e}}_{\alpha}(\mathbf{k})\cdot\mathbf{E}(\omega)=(\hat{\mathbf{e}}\cdot\hat{\mathbf{e}}_{\alpha}(\mathbf{k}))E(\omega)$. Fourier transforming Eq.~(\ref{eq:eom}) and projecting onto branch $\alpha$ give the mode equation
\begin{equation}
\label{eq:mode}
\bigl(-\rho\omega^{2}+C_{\alpha}(\omega)k^{2}\bigr)u_{\alpha}(\mathbf{k},\omega)=\Delta q(\mathbf{k})E_{\alpha}(\omega),
\end{equation}
with $k=|\mathbf{k}|$, $C_{L}(\omega)=M(\omega)=K+\frac{4}{3}G(\omega)$, and $C_{T_1}(\omega)=C_{T_2}(\omega)=G(\omega)$.

The field-parallel macroscopic polarization is
\begin{equation}
\label{eq:main-polarization-definition}
P(\omega)=\frac{1}{V}\int d^{3}r\,\Delta q(\mathbf{r})\,\hat{\mathbf{e}}\cdot\mathbf{u}(\mathbf{r},\omega).
\end{equation}
Combining this definition with the mode response obtained from Eq.~\eqref{eq:mode} gives the mode-sum form
\begin{equation}
\label{eq:main-polarization-kspace}
\frac{P(\omega)}{E(\omega)}
=\frac{1}{V}\sum_{\mathbf{k},\alpha}
\frac{|\Delta q(\mathbf{k})|^{2}
\bigl(\hat{\mathbf{e}}\cdot\hat{\mathbf{e}}_{\alpha}(\mathbf{k})\bigr)^{2}}
{-\rho\omega^{2}+C_{\alpha}(\omega)k^{2}} .
\end{equation}
We assume statistically homogeneous and isotropic IR-effective charge fluctuations.
With the overbar denoting a disorder average, this gives
\begin{equation}
\label{eq:main-disorder-correlation}
\overline{\Delta q(\mathbf{k})}=0,
\qquad
\overline{\Delta q(\mathbf{k})\Delta q(\mathbf{k}')^{*}}
=\delta_{\mathbf{k},\mathbf{k}'}\left\langle|\Delta q(k)|^{2}\right\rangle_{\rm dis},
\end{equation}
where $\left\langle|\Delta q(k)|^{2}\right\rangle_{\rm dis}$ depends only on $k=|\mathbf{k}|$. We also define the orientational factor
$\gamma_{\alpha}=\left\langle(\hat{\mathbf{e}}\cdot\hat{\mathbf{e}}_{\alpha})^{2}\right\rangle_{\rm angles}$.
After the disorder and orientational averages,
\begin{equation}
\label{eq:main-polarization-averaged}
\frac{P(\omega)}{E(\omega)}
=\frac{1}{V}\sum_{\mathbf{k},\alpha}
\frac{\left\langle|\Delta q(k)|^{2}\right\rangle_{\rm dis}\gamma_{\alpha}}
{-\rho\omega^{2}+C_{\alpha}(\omega)k^{2}} .
\end{equation}
Finally, replacing the discrete sum by the isotropic Debye-sphere form
\begin{equation}
\label{eq:main-debye-sphere}
\frac{1}{V}\sum_{\mathbf{k}}f(k)\rightarrow\int_{0}^{k_{\mathrm{D}}}f(k)\frac{3k^{2}}{k_{\mathrm{D}}^{3}}\,dk
\end{equation}
and using $\varepsilon(\omega)=\varepsilon_{\infty}+P(\omega)/[\varepsilon_{0}E(\omega)]$ gives
\begin{equation}
\label{eq:eps_sum}
\varepsilon(\omega)=\varepsilon_{\infty}+\frac{1}{\rho\varepsilon_{0}}\sum_{\alpha}\int_{0}^{k_{\mathrm{D}}}\frac{\left\langle|\Delta q(k)|^{2}\right\rangle_{\rm dis}\gamma_{\alpha}}{-\omega^{2}+\tilde{C}_{\alpha}(\omega)k^{2}}\frac{3k^{2}}{k_{\mathrm{D}}^{3}}\,dk,
\end{equation}
where $\tilde{C}_{\alpha}(\omega)=C_{\alpha}(\omega)/\rho$. Appendices~\ref{app:continuum-mode}--\ref{app:isotropic-limit} give the corresponding algebra and the branch-resolved form. In an isotropic three-dimensional elastic medium, the three acoustic polarizations form an orthonormal triad and are statistically equivalent, so that $\gamma_{L}=\gamma_{T_1}=\gamma_{T_2}=1/3$. Grouping the two degenerate transverse branches and assuming that $\left\langle|\Delta q(k)|^{2}\right\rangle_{\rm dis}$ is branch-independent, Eq.~(\ref{eq:eps_sum}) reduces to
\begin{equation}
\label{eq:eps_reduced}
\begin{aligned}
\varepsilon(\omega)=\varepsilon_{\infty}
+\frac{1}{\rho\varepsilon_{0}}
\int_{0}^{k_{\mathrm{D}}}
\Biggl[
&\frac{2}{3}\,
\frac{\left\langle|\Delta q(k)|^{2}\right\rangle_{\rm dis}}
{-\omega^{2}+\tilde{G}(\omega)k^{2}}
\\
&+\frac{1}{3}\,
\frac{\left\langle|\Delta q(k)|^{2}\right\rangle_{\rm dis}}
{-\omega^{2}+\tilde{M}(\omega)k^{2}}
\Biggr]
\frac{3k^{2}}{k_{\mathrm{D}}^{3}}\,dk,
\end{aligned}
\end{equation}
where $\tilde{G}(\omega)=G(\omega)/\rho$ and $\tilde{M}(\omega)=M(\omega)/\rho$, with $M(\omega)=K+(4/3)G(\omega)$.
Equation~(\ref{eq:eps_reduced}) is the expression used to analyze the THz dielectric response and the BP-related excess contribution.
It should be regarded as a factorized effective-medium closure: the IR-effective charge-coupling spectrum $\left\langle|\Delta q(k)|^{2}\right\rangle_{\rm dis}$ is introduced independently of the effective mechanical response specified by $G(\omega)$ and the bulk modulus $K$.
Possible connected correlations between local charge-coupling environments and the microscopic mechanical disorder underlying the BP-related response are therefore not explicitly included in this closure.
Such correlations would generate non-factorizable corrections to Eq.~(\ref{eq:eps_reduced}) and are beyond the scope of the present minimal model.

\section{Application to glycerol glass}

\begin{figure}[t]
  \centering
  \IfFileExists{fig-dielectric-response.pdf}{%
    \includegraphics[width=0.95\columnwidth]{fig-dielectric-response}%
  }{%
    \fbox{\parbox[c][0.30\columnwidth][c]{\columnwidth}{\centering 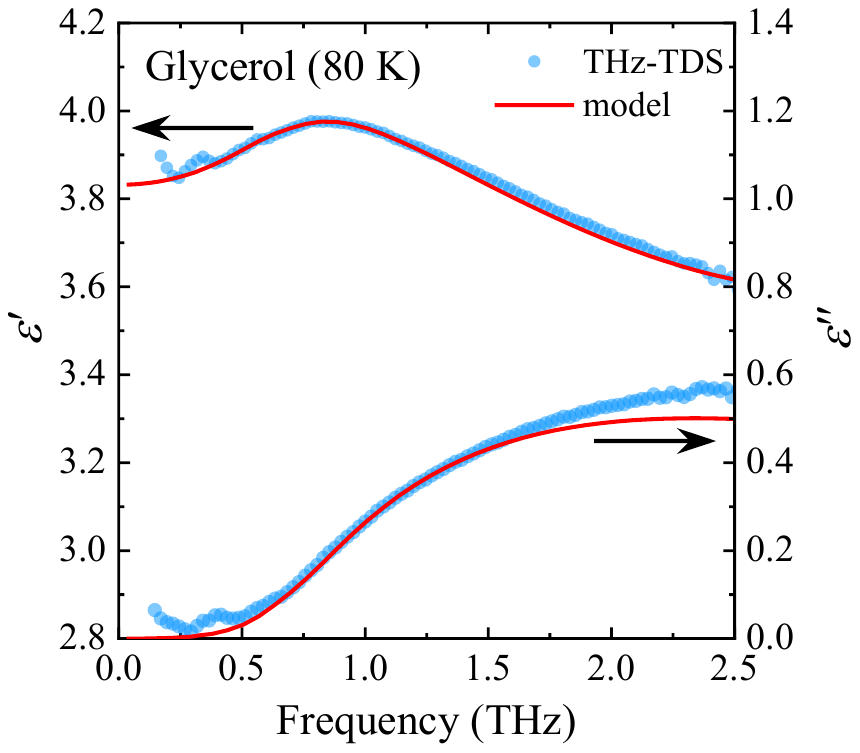 (placeholder)}}%
  }
  \caption{Complex dielectric response of glycerol glass at $80~\mathrm{K}$. Symbols: THz-TDS measurements; solid curves: model. Plotted are $\varepsilon'(\omega)$ and $\varepsilon''(\omega)$ versus frequency (THz). The model reproduces the crossover from resonance-like behavior below $\omega_{\rm BP}$ to a broad Debye-like response above $\omega_{\rm BP}$. The weak convex-upward feature in $\varepsilon'(\omega)$ near $\omega_{\rm BP}$ is consistent with a shallow dip in $G'(\omega)$, i.e., a reduction of $V_{\rm TA}(\omega)=\sqrt{G'(\omega)/\rho}$ associated with the BP.}
  \label{fig:dielectric-response}
\end{figure}

\nocite{Blieck2005,Chelli1999a,Chelli1999b,Egorov2011,Lorch1969,Faber-Ziman,Pasquarello1997,Pedesseau2015}
For quantitative comparison with experiment, the model inputs are $\rho$, $k_{\mathrm{D}}$, $G(\omega)$, $K$, and $\Delta q(k)$. For glycerol we use an effective shear modulus $G(\omega)$ obtained by applying the heterogeneous-elasticity-theory coherent-potential-approximation (HET-CPA) procedure described in Ref.~\cite{Kyotani2025} to the experimental vibrational density of states $g(\omega)$ of glycerol glass at $80~\mathrm{K}$, and parameterize the IR-effective charge spectrum by retaining the leading terms of a Maclaurin expansion, $\Delta q(k)\approx q_{0}+q_{2}k^{2}$.
The constant term $q_0$ is weakly constrained by the glycerol dielectric spectra: when $q_0$ and $q_2$ are varied together, the fitted $q_0$ remains consistent with zero within the fitting uncertainty.
In the main analysis we use the minimal form $\Delta q(k)=q_2 k^2$.
The coefficient $q_2$ is selected by a parameter scan to minimize the misfit between calculated and measured $\varepsilon'(\omega)$ and $\varepsilon''(\omega)$.
Details of the $q_0$ uncertainty and the $q_2$ scan are given in Appendix~\ref{app:parameter-scan}.
The parameters used in Fig.~\ref{fig:dielectric-response} are summarized in Table~\ref{tab:params}.

The complex dielectric function $\varepsilon(\omega)$ of glycerol glass was measured by terahertz time-domain
spectroscopy (THz-TDS)~\cite{Kabeya2016,Mori2020} using a commercial system (RT-10000, Tochigi Nikon Co.).
The usable frequency range in the present THz-TDS measurements was approximately $0.2$--$2.5~\mathrm{THz}$.
The THz beam path was enclosed in dry air, and the sample was mounted in a
liquid-helium flow cryostat for temperature control between $14$ and $295~\mathrm{K}$.

Glycerol ($\ge 99.5\%$, Sigma-Aldrich) was sealed in a liquid cell with a $0.2~\mathrm{mm}$ optical thickness between
two $z$-cut $\alpha$-quartz windows. The sample was vitrified \textit{in situ} by cooling through the glass transition
($T_g\simeq 190~\mathrm{K}$~\cite{Ryabov2003}); the data analyzed correspond to the glassy state at $80~\mathrm{K}$ after
equilibration. Optical constants were extracted from a standard slab-transmission analysis to obtain the complex
refractive index $n(\omega)$, and the complex dielectric function was then calculated via
$\varepsilon(\omega)=n^{2}(\omega)$.

\begin{figure}[t]
  \centering
    \IfFileExists{fig-shear-modulus-vdos.pdf}{%
    \includegraphics[width=\linewidth]{fig-shear-modulus-vdos}%
  }{%
    \fbox{%
      \parbox[c][0.48\textheight][c]{0.95\linewidth}{\centering
        {\small Placeholder for Fig.~\ref{fig:shear-modulus-vdos} (upload \texttt{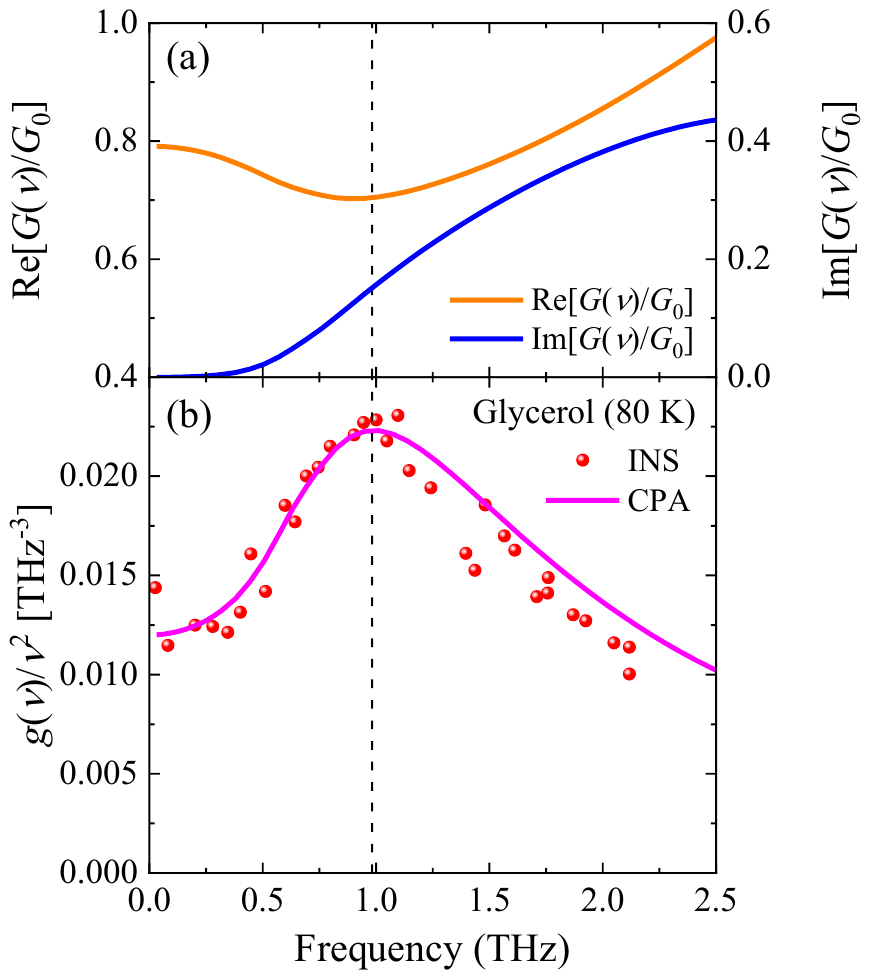} later)}
      }%
    }%
  }

  \caption{(a) Real and imaginary parts of the effective shear modulus $G(\omega)$ from the HET-CPA analysis,
  plotted as $\mathrm{Re}[G(\omega)/G_{0}]$ (orange) and $\mathrm{Im}[G(\omega)/G_{0}]$ (blue), with $G_0=5.43~\mathrm{GPa}$.
  (b) Reduced vibrational density of states $g(\omega)/\omega^{2}$ computed from $G(\omega)$
  (magenta line) compared with INS data~\cite{Wuttke1995} (filled circles). Here $\nu$ denotes the linear frequency, $\nu=\omega/2\pi$.}
  \label{fig:shear-modulus-vdos}
\end{figure}

\begin{table}[t]
\caption{\label{tab:params}
Parameters used to calculate the dielectric response of glycerol glass at $80~\mathrm{K}$ in Fig.~\ref{fig:dielectric-response}.
$\rho$ is the mass density, $k_{\mathrm{D}}$ the Debye wavenumber, $K$ the bulk modulus, and $\varepsilon_{\infty}$ the high-frequency dielectric constant.
$q_{0}$ and $q_{2}$ specify the quadratic parameterization of the IR-effective charge spectrum
$\Delta q(k)=q_{0}+q_{2}k^{2}$.
The fixed mechanical inputs and the HET-CPA parameter set defining the frequency-dependent shear modulus $G(\omega)$ are described in Appendix~\ref{app:shear-modulus}; the sound-velocity inputs are from Ref.~\cite{Scarponi2004}, and $G(\omega)$ is shown in Fig.~\ref{fig:shear-modulus-vdos}.
}
\squeezetable
\setlength{\tabcolsep}{1.3pt}
\begin{ruledtabular}
\begin{tabular}{lcccccc}
& \begin{tabular}[c]{@{}c@{}}$\rho$\\(g\,cm$^{-3}$)\end{tabular}
& \begin{tabular}[c]{@{}c@{}}$k_{\mathrm{D}}$\\(\AA$^{-1}$)\end{tabular}
& \begin{tabular}[c]{@{}c@{}}$K$\\(GPa)\end{tabular}
& \begin{tabular}[c]{@{}c@{}}$\varepsilon_{\infty}$\\\end{tabular}
& \begin{tabular}[c]{@{}c@{}}$q_{0}$\\(C\,cm$^{-3}$)\end{tabular}
& \begin{tabular}[c]{@{}c@{}}$q_{2}$\\(C\,cm$^{-3}$\,\AA$^{2}$)\end{tabular}
\\
\hline
Glycerol & 1.26 & 1.90 & 10.58 & 2.765 & 0 & $1.6\times10^{3}$ \\
\end{tabular}
\end{ruledtabular}
\end{table}

Figure~\ref{fig:dielectric-response} compares the measured complex dielectric response of glycerol glass with the model.
Over $0.3$--$2.5~\mathrm{THz}$, the model simultaneously reproduces both the real and imaginary parts,
$\varepsilon'(\omega)$ and $\varepsilon''(\omega)$.
Around the BP frequency $\omega_{\rm BP}$ the response crosses over from resonance-like behavior
below $\omega_{\rm BP}$ to a broad, Debye-like lineshape above $\omega_{\rm BP}$, in quantitative
agreement with experiment.
This agreement is naturally explained if optical attenuation is mediated by the effective complex shear modulus $G(\omega)$: Rayleigh-like scattering of
long-wavelength acoustic modes prevails below $\omega_{\rm BP}$, whereas above $\omega_{\rm BP}$
the dissipation associated with effective elastic heterogeneities is captured by a rapid increase
of the viscoelastic loss modulus $G''(\omega)$, which in turn produces the strong IR absorption and
broad Debye-like lineshape.
The small convex-upward bump in $\varepsilon'(\omega)$ near
$\omega_{\rm BP}$ is accounted for by a shallow dip in the storage modulus $G'(\omega)$ (Fig.~\ref{fig:shear-modulus-vdos}), which lowers the
transverse sound speed $V_{\rm TA}(\omega)=\sqrt{G'(\omega)/\rho}$, and slightly flattens the
$\omega$--$k$ dispersion, yielding an excess VDOS in the BP region; accordingly, the
$\varepsilon'(\omega)$ bump reflects the BP-related excess. Within our isotropic three-dimensional
formulation, Eq.~(\ref{eq:eps_reduced}), the transverse part of $\varepsilon(\omega)$ is governed
directly by the shear modulus $G(\omega)$, whereas the longitudinal part involves the longitudinal
modulus $M(\omega)=K+\frac{4}{3}G(\omega)$. In the present model, $K$ is a real,
frequency-independent constant; thus all frequency dependence and dissipation of $M(\omega)$
originate from $G(\omega)$, while the constant $K$ raises the longitudinal stiffness.
A decomposition of the modeled dielectric function into
transverse and longitudinal contributions (Fig.~\ref{fig:dielectric-decomposition}) shows that, in the BP region, both
$\varepsilon'(\omega)$ and $\varepsilon''(\omega)$ are almost entirely controlled by the transverse
channel, with the longitudinal contribution remaining comparatively small over the whole band.
Thus the BP-related THz dielectric response is controlled primarily by the frequency dependence
of $G(\omega)$ and the associated transverse-like vibrations.
At the same time, a fully microscopic derivation of the dielectric response remains a challenge.

\begin{figure}[t]
  \centering
  \IfFileExists{fig-reduced-spectra-ir-coupling.pdf}{%
    \includegraphics[width=0.90\columnwidth]{fig-reduced-spectra-ir-coupling}%
  }{%
    \fbox{\parbox[c][0.35\columnwidth][c]{\columnwidth}{\centering 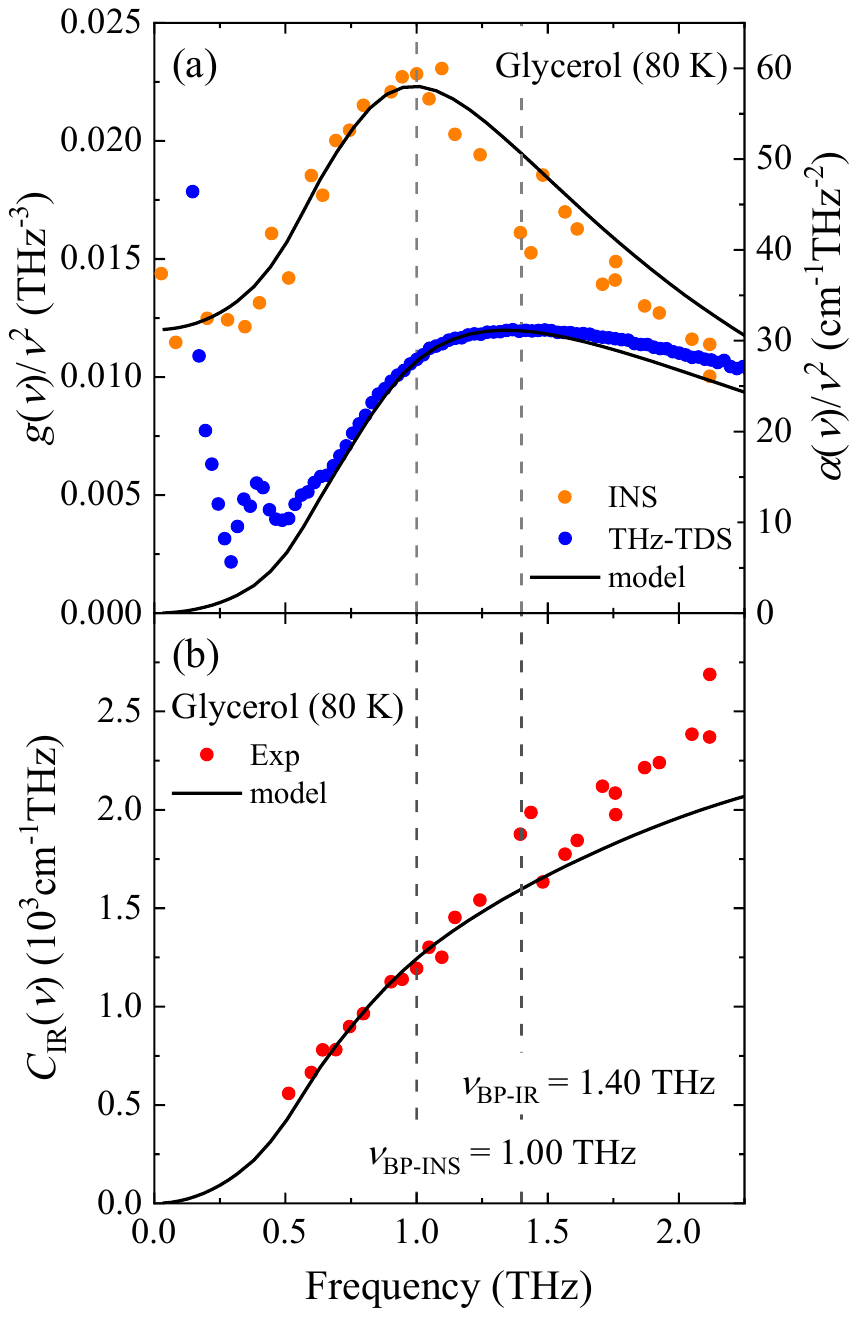 (placeholder)}}%
  }
  \caption{Reduced spectra and IR light-vibration coupling. (a) Reduced absorption $\alpha(\omega)/\omega^{2}$ (blue symbols) obtained from Fig.~\ref{fig:dielectric-response} via $\alpha(\omega)=\omega\varepsilon''(\omega)/(c\,n'(\omega))$ (with $c$ the speed of light and $n'(\omega)$ the real part of the refractive index), together with the reduced VDOS $g(\omega)/\omega^{2}$ (orange symbols) from independent inelastic neutron scattering (INS)~\cite{Wuttke1995}; solid lines are the model. (b) $C_{\rm IR}(\omega)=\alpha(\omega)/g(\omega)$. The near-linear dependence around the BP is captured by the model, whereas the quadratic Taraskin form $A+B\omega^{2}$ does not capture this trend. Vertical dashed lines mark the BP positions from IR and INS, $\omega_{\rm BP\text{-}IR}/2\pi \approx 1.40~\mathrm{THz}$ and $\omega_{\rm BP\text{-}INS}/2\pi \approx 1.00~\mathrm{THz}$, respectively.}
  \label{fig:reduced-spectra-ir-coupling}
\end{figure}

Using the modeled dielectric function, we convert $\varepsilon''(\omega)$ to the absorption coefficient
$\alpha(\omega)$ and, from the same effective modulus $G(\omega)$, obtain $g(\omega)$; the reduced spectra
$\alpha(\omega)/\omega^{2}$ and $g(\omega)/\omega^{2}$ are plotted in Fig.~\ref{fig:reduced-spectra-ir-coupling}(a). We then form
$C_{\rm IR}(\omega)=\alpha(\omega)/g(\omega)$ and compare with experiment. As shown in Fig.~\ref{fig:reduced-spectra-ir-coupling}(b), the
resulting $C_{\rm IR}(\omega)$ agrees with the data in the vicinity of the BP and captures the near-linear
trend there, whereas the Taraskin form $A+B\omega^{2}$~\cite{Taraskin2006} cannot account for this trend.

\section{Static charge correlations and coupling spectrum}

In Eq.~(\ref{eq:eps_reduced}) 
all microscopic information on the THz (IR) charge coupling is collected into the
$k$-dependent oscillator strength $\langle|\Delta q(k)|^2\rangle_{\rm dis}$.
Since the $k$ integral is restricted to the acoustic window $0<k<k_{\mathrm{D}}$, we only require
$\Delta q(k)$ over this range and adopt the minimal analytic form
$\Delta q(k) \approx q_0+q_2 k^2$.
Microscopically, the mode oscillator strength is set by the mode effective charge $\mathbf{F}^{(p)}$~\cite{Thorpe1986}, i.e., the projection of the rigid-ion charge distribution onto each eigenmode $p$ (equivalently by the charge--charge dynamic structure factor $S_{ZZ}(k,\omega)$ and its projection onto the eigenmodes~\cite[Chap.~10]{Hansen2006}).
Appendix~\ref{app:microscopic-szz} relates this mode-resolved quantity to static charge correlations in a limited sense.
There, disordered eigenmodes are projected onto longitudinal and transverse plane-wave components with mode-dependent $k$-space weights $a_{\mathbf{k}\alpha}(\omega_p)$~\cite{Taraskin2000,Taraskin2006}.
Before further approximation, the squared projected mode charge contains the configurational four-point average
$\langle a_{\mathbf{k}\alpha}(\omega_p)a^{*}_{\mathbf{k}'\alpha'}(\omega_p)\rho_Z(\mathbf{k})\rho_Z(\mathbf{k}')^{*}\rangle_{\rm conf}$.
Under a diagonal, factorized approximation---neglecting cross/interference terms and representing the remaining charge correlations by the static structure factor $S_{ZZ}(k)$---this gives
$\left\langle\,|\hat{\mathbf{e}}\cdot\mathbf{F}^{(p)}|^2\,\right\rangle \propto \sum_{\mathbf{k},\alpha}|a_{\mathbf{k}\alpha}(\omega_p)|^2 S_{ZZ}(k)\,\gamma_{\alpha}$.
This construction provides a microscopic route and diagnostic by which a static $S_{ZZ}(k)$ can contribute to the effective coupling spectrum represented by $\langle|\Delta q(k)|^2\rangle_{\rm dis}$ in Eq.~(\ref{eq:eps_reduced}), rather than a full microscopic determination of that spectrum.
This diagonal/factorized interpretation is expected to be most transparent in the low-frequency propagon regime, where the $k$-space weights are relatively narrow; above the BP, the weights broaden as the modes become diffuson-like, so the neglected off-diagonal terms and possible connected correlations may become more important.
A fully microscopic determination of this effective charge-coupling spectrum would require dynamic, mode-resolved charge-correlation information, such as $S_{ZZ}(k,\omega)$ or an equivalent eigenmode-resolved effective-charge analysis, but the static $S_{ZZ}(k)$ from molecular-dynamics (MD) suffices to diagnose whether medium-range correlations (e.g., the first sharp diffraction peak (FSDP)) should appear within the relevant acoustic window.

\begin{figure}[t]
  \centering
  \IfFileExists{fig-static-structure-factors.pdf}{%
    \includegraphics[width=0.90\columnwidth]{fig-static-structure-factors}%
  }{%
    \fbox{\parbox[c][0.35\columnwidth][c]{\columnwidth}{\centering 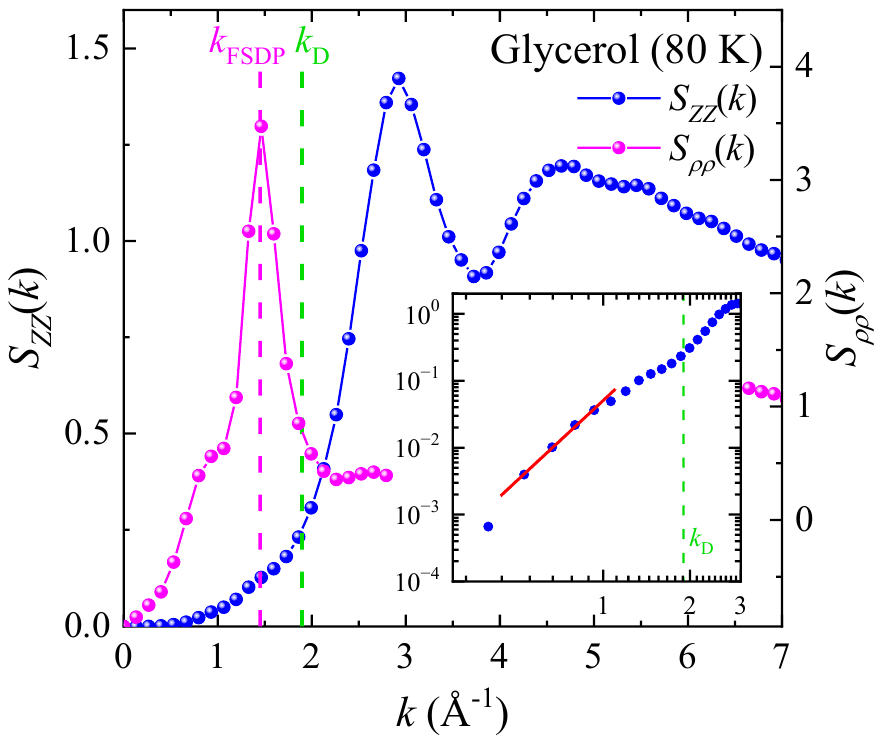 (placeholder)}}%
  }
  \caption{Static charge and density correlations in glycerol glass ($80~\mathrm{K}$). Shown are the charge--charge structure factor $S_{ZZ}(k)$ (blue) and the mass-density static structure factor $S_{\rho\rho}(k)$ (magenta) obtained from MD simulations. The FSDP and Debye wavenumbers ($k_{\rm FSDP}$, $k_{\mathrm{D}}$) are indicated. $S_{\rho\rho}(k)$ exhibits a pronounced FSDP at $k \sim 1.5~\text{\AA}^{-1}$, whereas $S_{ZZ}(k)$ has no FSDP maximum; instead its first (lowest-$k$) maximum appears only at higher $k$, on the interatomic-spacing scale. The inset shows $S_{ZZ}(k)$ on double-logarithmic axes in the low-$k$ region. A power-law fit $S_{ZZ}(k)=A k^{n}$ over the lowest numerically reliable interval $0.532\leq k\leq0.931~\text{\AA}^{-1}$ gives $n\simeq4.0$, with coefficient of determination $R_{\mathrm{fit}}^{2}=0.998$. The red line shows this fit, extended beyond the fitting interval as a guide to the eye. The fitted dependence is consistent with the $k^{4}$ oscillator-strength scaling implied by the minimal form $\Delta q(k)=q_{2}k^{2}$.}
  \label{fig:static-structure-factors}
\end{figure}

For glycerol, $S_{ZZ}(k)$ exhibits no maximum at the FSDP ($k\simeq 1.5\,\text{\AA}^{-1}$~\cite{Champeney1986}) and varies smoothly throughout the acoustic window (Fig.~\ref{fig:static-structure-factors}).
To examine whether the adopted minimal form $\Delta q(k)=q_{2}k^{2}$ is compatible with the MD charge correlations, the inset of Fig.~\ref{fig:static-structure-factors} shows $S_{ZZ}(k)$ on double-logarithmic axes.
Because the smallest-$k$ values are affected by the finite simulation cell and the real-space truncation used in evaluating the pair-correlation integrals, we restrict the comparison to the lowest numerically reliable interval, $0.532\leq k\leq0.931~\text{\AA}^{-1}$.
A free power-law fit $S_{ZZ}(k)=A k^{n}$ over this interval gives $n\simeq4.0$, with coefficient of determination $R_{\mathrm{fit}}^{2}=0.998$, showing that the accessible low-$k$ behavior is compatible with a $k^{4}$ dependence.

Within the limited diagonal and factorized connection established in Appendix~\ref{app:microscopic-szz}, this agreement provides a structural consistency check for the oscillator-strength dependence $\langle|\Delta q(k)|^{2}\rangle_{\rm dis}\propto k^{4}$ implied by $\Delta q(k)=q_{2}k^{2}$.
The absence of an FSDP maximum likewise provides no motivation for adding an explicit FSDP-like structure to $\Delta q(k)$.
Independently, the dielectric fit yields a value of $q_{0}$ consistent with zero, as quantified in Appendix~\ref{app:parameter-scan}.
Accordingly, for Fig.~\ref{fig:dielectric-response} we use the minimal form $\Delta q(k)=q_{2}k^{2}$ and select $q_{2}$ by the parameter scan described above.
With this choice the model yields a near-linear $C_{\rm IR}(\omega)$ in the vicinity of the BP.

\begin{figure}[t]
  \centering
  \IfFileExists{fig-coupling-ratio-scan.pdf}{%
    \includegraphics[width=\columnwidth]{fig-coupling-ratio-scan}%
  }{%
    \fbox{\parbox[c][0.35\columnwidth][c]{\columnwidth}{\centering 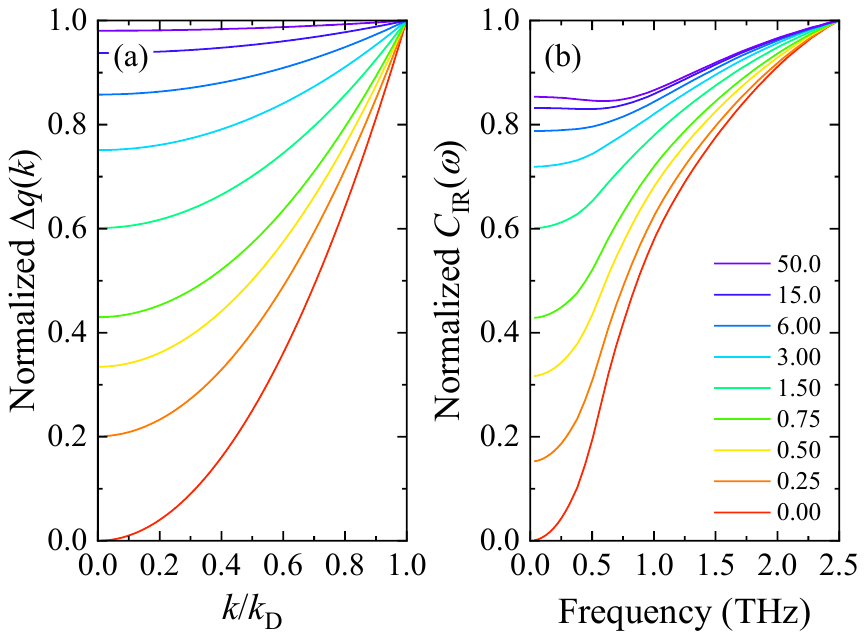 (placeholder)}}%
  }
  \caption{Evolution with $R=q_{0}/(q_{2}k_{\mathrm{D}}^{2})$ in $\Delta q(k)=q_{0}+q_{2}k^{2}$: approaching the Taraskin-like regime.
  For each $R$, $q_0$ and $q_2$ are rescaled to keep $\int_{0}^{k_{\mathrm{D}}} \! dk\, k^{2}\lvert \Delta q(k)\rvert^{2}$ constant, isolating the effect of the $k$-independent term $q_0$ (see Appendix~\ref{app:r-scan}).
  (a) Normalized $\Delta q(k)$ (scaled to unity at $k=k_{\mathrm{D}}$) plotted as a function of $k/k_{\mathrm{D}}$ for several $R$.
  (b) Normalized $C_{\rm IR}(\omega)$ (scaled to unity at $2.5~\mathrm{THz}$) computed with the same $G(\omega)$. Increasing $R$ strengthens the $k$-independent part of $\Delta q(k)$ and drives $C_{\mathrm{IR}}(\omega)$ from a nearly linear form (small $R$) toward the Taraskin-like behavior $C_{\mathrm{IR}}(\omega)\simeq A + B\omega^{2}$.}
  \label{fig:coupling-ratio-scan}
\end{figure}

We next examine how this near-linear behavior changes when a finite constant term $q_{0}$ is introduced. To this end, we vary the dimensionless ratio $R=q_{0}/(q_{2}k_{\mathrm{D}}^{2})$, which sets the balance between the $k$-independent ($q_{0}$) and quadratic ($q_{2}k^{2}$) parts in $\Delta q(k)$.
Increasing $q_{0}$ enhances the $k$-independent weight of
$\langle|\Delta q(k)|^{2}\rangle_{\rm dis}$---white in $k$---which, upon Fourier synthesis with random phases
over $k<k_{\mathrm{D}}$, corresponds to essentially white (spatially uncorrelated) charge fluctuations in real
space. As the ratio $R$ increases (Fig.~\ref{fig:coupling-ratio-scan}(a)), the computed $C_{\rm IR}(\omega)$ evolves continuously
toward a constant-term-dominated form (Fig.~\ref{fig:coupling-ratio-scan}(b)).
Within our framework the mechanism is direct: in the limit where the $k$-independent component
dominates the $\varepsilon(\omega)$ integral ($\Delta q(k)\approx q_{0}$), the absorption $\alpha(\omega)$ nearly follows the VDOS, $\alpha(\omega)\propto g(\omega)/n'(\omega)$; hence $C_{\mathrm{IR}}(\omega)$ becomes only weakly frequency dependent.

\section{Relation to the Taraskin coupling form}

Finally, we turn to the question of why Taraskin's model does not reproduce the nearly linear
$C_{\rm IR}(\omega)$ observed near the BP. Taraskin's approach starts from a factorization ansatz
$\alpha(\omega)\approx C_{\rm IR}(\omega)g(\omega)$~\cite{Taraskin2006}, motivated by the mode-resolved linear-response expression for crystals derived by Maradudin~\cite{Maradudin1961}.
Microscopically, however, linear response gives $\alpha(\omega)=\sum_{p} C_{p}\delta(\omega-\omega_{p})$, i.e., a sum over modes with mode-dependent couplings $C_{p}$, not a product of a smooth coupling and the VDOS
$g(\omega)=\sum_{p}\delta(\omega-\omega_{p})$. Identifying $\alpha(\omega)$ with
$C_{\rm IR}(\omega)g(\omega)$ amounts to a modeling assumption: it replaces the mode-resolved couplings $C_{p}$
by a single smooth function of frequency $C_{\rm IR}(\omega)$.
This limitation was emphasized early on for disordered solids, where absorption involves mode- (or subband-) resolved couplings that can vary strongly with frequency (see Galeener and Sen~\cite{Galeener1978}).
A related non-factorizable convolution form appears in low-frequency Raman scattering (Schmid and Schirmacher~\cite{Schmid2008}).
Under Taraskin's decomposition into uncorrelated and correlated charge components, this factorization yields the quadratic form $C_{\rm IR}(\omega)=A+B\omega^{2}$ used in their analysis.
Within our framework, Eq.~(\ref{eq:eps_reduced}) can be interpreted as an effective continuum $k$-space representation of
Maradudin's mode-resolved linear-response formula~\cite{Maradudin1961},
in which the discrete mode sum is replaced by an integral over $k$ and
the mode oscillator strengths are represented by the $k$-dependent spectrum $\langle|\Delta q(k)|^{2}\rangle_{\rm dis}$ (see Appendices~\ref{app:continuum-mode}--\ref{app:isotropic-limit} and Appendix~\ref{app:microscopic-szz}).
In this effective representation, the spectral factor is the viscoelastic Green's function of the acoustic branches,
$\left[-\omega^{2}+\tilde{C}_{\alpha}(\omega)k^{2}\right]^{-1}$, determined by the complex moduli $G(\omega)$ and
$M(\omega)$. A fully microscopic determination of these effective inputs is left for future work.

\section{Conclusions}

A continuum model based on IR-effective charge fluctuations $\Delta q(k)$ and a
frequency-dependent shear modulus $G(\omega)$ reproduces the measured $\varepsilon'(\omega)$,
$\varepsilon''(\omega)$, and the near-linear $C_{\rm IR}(\omega)$ around the BP in glycerol glass.
The single ratio $R=q_{0}/(q_{2}k_{\mathrm{D}}^{2})$ tunes the IR coupling from this linear regime toward the
constant/quadratic-coupling behavior widely used to describe THz absorption in disordered materials,
$C_{\rm IR}(\omega)=A+B\omega^{2}$.
With material-specific inputs, the framework is readily extendable to other glasses and provides a
practical route to assess THz dielectric losses.


\begin{acknowledgments}
This work was supported by JSPS KAKENHI Grant Nos.~23H01139 and 23K25836 (to T.M.),
25H01519 and 22K03543 (to H.M.), and 24K08045 (to Y.F.);
by the Asahi Glass Foundation (to T.M.);
and by support from GIC \& NGF (to T.M.).
\end{acknowledgments}

\appendix
\twocolumngrid

\section{Microscopic origin of the electric driving term}
\label{app:electric-driving}

Here we derive the electric source term in Eq.~\eqref{eq:eom} from the microscopic rigid-ion dipole.
The mechanical part of Eq.~\eqref{eq:eom} is represented in the continuum description by the elastic bulk modulus $K$ and the effective, generally dissipative shear modulus $G(\omega)$.

We first work in the time domain.
The external electric field is assumed to be spatially uniform over the length scale of the vibrational displacement field.
When a polarization direction is specified, we write
\begin{equation}
\label{eq:appA_field_time}
\mathbf{E}(t)=E(t)\hat{\mathbf{e}},
\end{equation}
where $\hat{\mathbf{e}}$ is a time-independent unit vector.
The corresponding frequency-domain convention is
$\mathbf{E}(\omega)=E(\omega)\hat{\mathbf{e}}$, as used in the main text.

For a microscopic configuration, the displaced position and the decomposition of the microscopic dipole into static and vibrational parts are given in Eqs.~\eqref{eq:main-position}--\eqref{eq:main-dvib-discrete} of the main text.
In the rigid-ion approximation the charges $q_i$ are fixed to the moving sites.
The static term $\mathbf{d}_{0}$ is independent of the vibrational coordinates and therefore does not contribute to the vibrational driving force, whereas $\mathbf{d}^{\rm vib}(t)$ is the dynamical part relevant for infrared absorption.

To pass to the continuum model, introduce the microscopic rigid-ion charge distribution in the reference configuration,
\begin{equation}
\label{eq:appA_qri}
q_{\rm ri}(\mathbf{r})=\sum_i q_i\,\delta(\mathbf{r}-\mathbf{R}_{i}^{0}),
\end{equation}
and identify the discrete displacement with the continuum displacement at the corresponding material point,
\begin{equation}
\label{eq:appA_ui_to_u}
\mathbf{u}_{i}(t)\simeq\mathbf{u}(\mathbf{R}_{i}^{0},t).
\end{equation}
Then the vector vibrational dipole can be written as
\begin{equation}
\label{eq:appA_dvib_integral_qri}
\mathbf{d}^{\rm vib}(t)
\simeq\int d^{3}r\,q_{\rm ri}(\mathbf{r})\,\mathbf{u}(\mathbf{r},t).
\end{equation}
In the continuum-field description used in the main text, the reference-configuration site distribution is represented by the local IR-effective charge fluctuation field $\Delta q(\mathbf{r})$,
\begin{equation}
\label{eq:appA_qri_to_dq}
q_{\rm ri}(\mathbf{r})\rightarrow\Delta q(\mathbf{r}).
\end{equation}
Applying this replacement to Eq.~\eqref{eq:appA_dvib_integral_qri} gives the continuum-field vibrational dipole in Eq.~\eqref{eq:main-dvib-continuum} of the main text.
The position vector $\mathbf{r}$ labels a material point in the equilibrium reference configuration, and the spatial dependence of the IR-effective charge fluctuations is retained in $\Delta q(\mathbf{r})$.
The time dependence of the vibrational dipole is carried by $\mathbf{u}(\mathbf{r},t)$, which represents the site displacements in Eq.~\eqref{eq:appA_ui_to_u}.
If one writes the instantaneous microscopic rigid-ion charge density in real space,
with $\mathbf{x}$ denoting a position vector in the instantaneous configuration,
\begin{equation}
\label{eq:appA_rho_inst}
\rho_{\rm ri}(\mathbf{x},t)
=\sum_i q_i\,\delta\!\left[\mathbf{x}-\mathbf{R}_{i}^{0}-\mathbf{u}_{i}(t)\right],
\end{equation}
that density describes the positions of the moving charged sites.
This should be distinguished from the reference-configuration field $\Delta q(\mathbf{r})$ in Eq.~\eqref{eq:main-dvib-continuum}, which multiplies the displacement field in the linear vibrational dipole.
For a neutral microscopic cell, the zero-wave-number component of the bare rigid-ion charge distribution $q_{\rm ri}(\mathbf{r})$ is constrained by charge neutrality.
The charge-coupling quantity entering the dielectric response below is the effective disorder variance $\langle |\Delta q(k)|^{2}\rangle_{\rm dis}$.

The interaction energy and the corresponding electric force density are given in Eqs.~\eqref{eq:main-interaction-energy} and \eqref{eq:main-electric-force-density} of the main text.
After Fourier transformation in time, the latter supplies the electric driving term $\Delta q(\mathbf{r})\mathbf{E}(\omega)$ in Eq.~\eqref{eq:eom}.

\section{Continuum equation and mode decomposition}\label{app:continuum-mode}
We start from the elastodynamic equation in Eq.~\eqref{eq:eom} and use the Fourier representation and the associated reality condition, acoustic polarization basis, and branch projection given in Eq.~\eqref{eq:main-fourier-representation} and the accompanying text.
Here $u_{\alpha}(\mathbf{k},\omega)$ is a complex response amplitude at a fixed frequency; the amplitude at $-\mathbf{k}$ is obtained by applying the same mode equation to the wave vector $-\mathbf{k}$.
With these definitions, Fourier transforming Eq.~\eqref{eq:eom} and projecting onto branch $\alpha$ gives the mode equation, Eq.~\eqref{eq:mode} of the main text.
Solving Eq.~\eqref{eq:mode} for $u_{\alpha}(\mathbf{k},\omega)$ gives
\begin{equation}\label{eq:appS11}
u_{\alpha}(\mathbf{k},\omega)=
\frac{\Delta q(\mathbf{k})}{-\rho \omega^{2}+C_{\alpha}(\omega)k^{2}}\,E_{\alpha}(\omega).
\end{equation}

\section{Macroscopic polarization and the appearance of $|\Delta q(\mathbf{k})|^2$}\label{app:polarization-dq2}
Starting from the field-parallel macroscopic polarization defined in Eq.~\eqref{eq:main-polarization-definition} and inserting the Fourier representation in Eq.~\eqref{eq:main-fourier-representation} yields
\begin{equation}\label{eq:appS13}
\begin{aligned}
P(\omega)
&=\frac{1}{V}\int d^{3}r
\left[\frac{1}{\sqrt{V}}\sum_{\mathbf{k}}
\Delta q(\mathbf{k})e^{\mathrm{i}\mathbf{k}\cdot\mathbf{r}}\right] \\
&\quad\times
\left[\frac{1}{\sqrt{V}}\sum_{\mathbf{k}',\alpha}u_{\alpha}(\mathbf{k}',\omega)
\bigl(\hat{\mathbf{e}}\cdot\hat{\mathbf{e}}_{\alpha}(\mathbf{k}')\bigr)
e^{\mathrm{i}\mathbf{k}'\cdot\mathbf{r}}\right] \\
&=\frac{1}{V}\sum_{\mathbf{k},\mathbf{k}',\alpha}
\Delta q(\mathbf{k})u_{\alpha}(\mathbf{k}',\omega)
\bigl(\hat{\mathbf{e}}\cdot\hat{\mathbf{e}}_{\alpha}(\mathbf{k}')\bigr) \\
&\quad\times
\int d^{3}r\,e^{\mathrm{i}(\mathbf{k}+\mathbf{k}')\cdot\mathbf{r}}.
\end{aligned}
\end{equation}
The spatial integral produces a Kronecker delta (for a discrete $\mathbf{k}$-mesh),
\begin{equation}\label{eq:appS14}
\int d^{3}r\,e^{\mathrm{i}(\mathbf{k}+\mathbf{k}')\cdot\mathbf{r}}
=V\delta_{\mathbf{k}',-\mathbf{k}},
\end{equation}
so that
\begin{equation}\label{eq:appS15}
P(\omega)=\frac{1}{V}\sum_{\mathbf{k},\alpha}\Delta q(\mathbf{k})u_{\alpha}(-\mathbf{k},\omega)
\bigl(\hat{\mathbf{e}}\cdot\hat{\mathbf{e}}_{\alpha}(-\mathbf{k})\bigr).
\end{equation}
The amplitude at $-\mathbf{k}$ follows by applying the mode equation, Eq.~\eqref{eq:mode}, to the wave vector $-\mathbf{k}$:
\begin{equation}\label{eq:appS16}
u_{\alpha}(-\mathbf{k},\omega)
=
\frac{\Delta q(-\mathbf{k})}{-\rho \omega^{2}+C_{\alpha}(\omega)k^{2}}\,
\bigl[\hat{\mathbf{e}}_{\alpha}(-\mathbf{k})\cdot\mathbf{E}(\omega)\bigr].
\end{equation}
Substituting Eq.~\eqref{eq:appS16} into Eq.~\eqref{eq:appS15} gives
\begin{equation}\label{eq:appS18}
\begin{aligned}
P(\omega)
&=\frac{E(\omega)}{V}\sum_{\mathbf{k},\alpha}
\frac{\Delta q(\mathbf{k})\Delta q(-\mathbf{k})}
{-\rho \omega^{2}+C_{\alpha}(\omega)k^{2}}
\bigl(\hat{\mathbf{e}}\cdot\hat{\mathbf{e}}_{\alpha}(-\mathbf{k})\bigr)^{2} \\
&=\frac{E(\omega)}{V}\sum_{\mathbf{k},\alpha}
\frac{|\Delta q(\mathbf{k})|^{2}}
{-\rho \omega^{2}+C_{\alpha}(\omega)k^{2}}
\bigl(\hat{\mathbf{e}}\cdot\hat{\mathbf{e}}_{\alpha}(\mathbf{k})\bigr)^{2}.
\end{aligned}
\end{equation}
In the second line we used the reality of the charge-coupling field in the reference configuration, $\Delta q(-\mathbf{k})=\Delta q(\mathbf{k})^{*}$.
The squared projection is unchanged by the possible sign change of $\hat{\mathbf e}_{\alpha}$ between $\mathbf{k}$ and $-\mathbf{k}$.
Thus the factor $|\Delta q(\mathbf{k})|^{2}$ arises from the convolution of the charge-coupling field at $\mathbf{k}$ with the displacement response at $-\mathbf{k}$, together with the reality condition for $\Delta q(\mathbf{r})$.
At this stage we have not yet made any statistical assumptions about $\Delta q(\mathbf{r})$ or about the
orientation of $\mathbf{k}$ and $\hat{\mathbf{e}}_{\alpha}$.

\section{Disorder average and orientational coupling factors}\label{app:disorder-average}
We use the statistically homogeneous and isotropic disorder correlation defined in Eq.~\eqref{eq:main-disorder-correlation} of the main text.
Averaging Eq.~\eqref{eq:appS18} over the disorder using that relation gives
\begin{equation}\label{eq:appS20}
\overline{P(\omega)}=\frac{E(\omega)}{V}\sum_{\mathbf{k},\alpha}
\frac{\langle|\Delta q(k)|^{2}\rangle_{\mathrm{dis}}}{-\rho \omega^{2}+C_{\alpha}(\omega)k^{2}}
(\hat{\mathbf{e}}\cdot\hat{\mathbf{e}}_{\alpha})^{2}.
\end{equation}
We now perform an orientational average over the relative angle between the external field direction
$\hat{\mathbf{e}}$ and the polarization basis $\{\hat{\mathbf{e}}_{\alpha}(\mathbf{k})\}$.
We assume that the statistics of $\Delta q(\mathbf{k})$ are independent of these orientations;
hence the disorder and orientational averages factorize. We define the orientational coupling factor
\begin{equation}\label{eq:appS21}
\gamma_{\alpha}\equiv \langle(\hat{\mathbf{e}}\cdot\hat{\mathbf{e}}_{\alpha})^{2}\rangle_{\mathrm{angles}},
\end{equation}
and replace $(\hat{\mathbf{e}}\cdot\hat{\mathbf{e}}_{\alpha})^{2}$ in Eq.~\eqref{eq:appS20} by its angular average, giving Eq.~\eqref{eq:main-polarization-averaged} of the main text.
Using $\varepsilon(\omega)=\varepsilon_{\infty}+\overline{P(\omega)}/[\varepsilon_{0}E(\omega)]$ with that result yields the general discrete-$\mathbf{k}$ expression for the dielectric function:
\begin{equation}\label{eq:appS23}
\varepsilon(\omega)=\varepsilon_{\infty}
+\frac{1}{\rho\varepsilon_{0}V}\sum_{\mathbf{k},\alpha}
\frac{\langle|\Delta q(k)|^{2}\rangle_{\mathrm{dis}}\,\gamma_{\alpha}}
{-\omega^{2}+\tilde{C}_{\alpha}(\omega)k^{2}},
\end{equation}
where $\tilde{C}_{\alpha}(\omega)=C_{\alpha}(\omega)/\rho$.
Next we apply the isotropic Debye-sphere replacement in Eq.~\eqref{eq:main-debye-sphere} of the main text to Eq.~\eqref{eq:appS23}.
This gives the general branch-resolved dielectric function in Eq.~\eqref{eq:eps_sum} of the main text.
The replacement follows the standard isotropic Debye prescription, whereby the discrete $\mathbf{k}$ sum is approximated by a continuum integral over the isotropic Debye sphere $0<k<k_{\mathrm{D}}$, with $k_{\mathrm{D}}$ providing an interatomic-scale (short-wavelength) cutoff for the present continuum formulation.

\section{Isotropic limit: $\gamma = 1/3$ and the final expression}\label{app:isotropic-limit}
In an isotropic three-dimensional elastic medium the three polarization vectors
$\{\hat{\mathbf{e}}_{L},\hat{\mathbf{e}}_{T_{1}},\hat{\mathbf{e}}_{T_{2}}\}$
form an orthonormal basis for any given $\mathbf{k}$. For a fixed $\hat{\mathbf{e}}$ we always have
\begin{equation}\label{eq:appS26}
(\hat{\mathbf{e}}\cdot\hat{\mathbf{e}}_{L})^{2}
+(\hat{\mathbf{e}}\cdot\hat{\mathbf{e}}_{T_{1}})^{2}
+(\hat{\mathbf{e}}\cdot\hat{\mathbf{e}}_{T_{2}})^{2}
=|\hat{\mathbf{e}}|^{2}=1.
\end{equation}
Because the three directions are statistically equivalent, their orientational averages are identical:
\begin{equation}\label{eq:appS27}
\gamma_{L}=\gamma_{T_{1}}=\gamma_{T_{2}}\equiv \gamma.
\end{equation}
Taking the orientational average of Eq.~\eqref{eq:appS26} immediately gives
\begin{equation}\label{eq:appS28}
3\gamma=1\ \Rightarrow\ \gamma=\frac{1}{3}.
\end{equation}
Substituting Eq.~\eqref{eq:appS28} into Eq.~\eqref{eq:eps_sum} and grouping the two degenerate transverse branches
($C_{T_{1}}=C_{T_{2}}=G$) and the single longitudinal branch ($C_{L}=M$) gives Eq.~\eqref{eq:eps_reduced} of the main text, with $\tilde{G}(\omega)$ and $\tilde{M}(\omega)$ defined immediately below that equation.

\section{Branch-resolved polarizations $P_T(\omega)$ and $P_L(\omega)$ and the dielectric contributions}\label{app:branch-decomposition}
\begin{figure}[t]
  \centering
    \IfFileExists{fig-dielectric-decomposition.pdf}{%
    \includegraphics[width=\linewidth]{fig-dielectric-decomposition}%
  }{%
    \fbox{%
      \parbox[c][0.48\textheight][c]{0.95\linewidth}{\centering
        {\small Placeholder for Fig.~\ref{fig:dielectric-decomposition} (upload \texttt{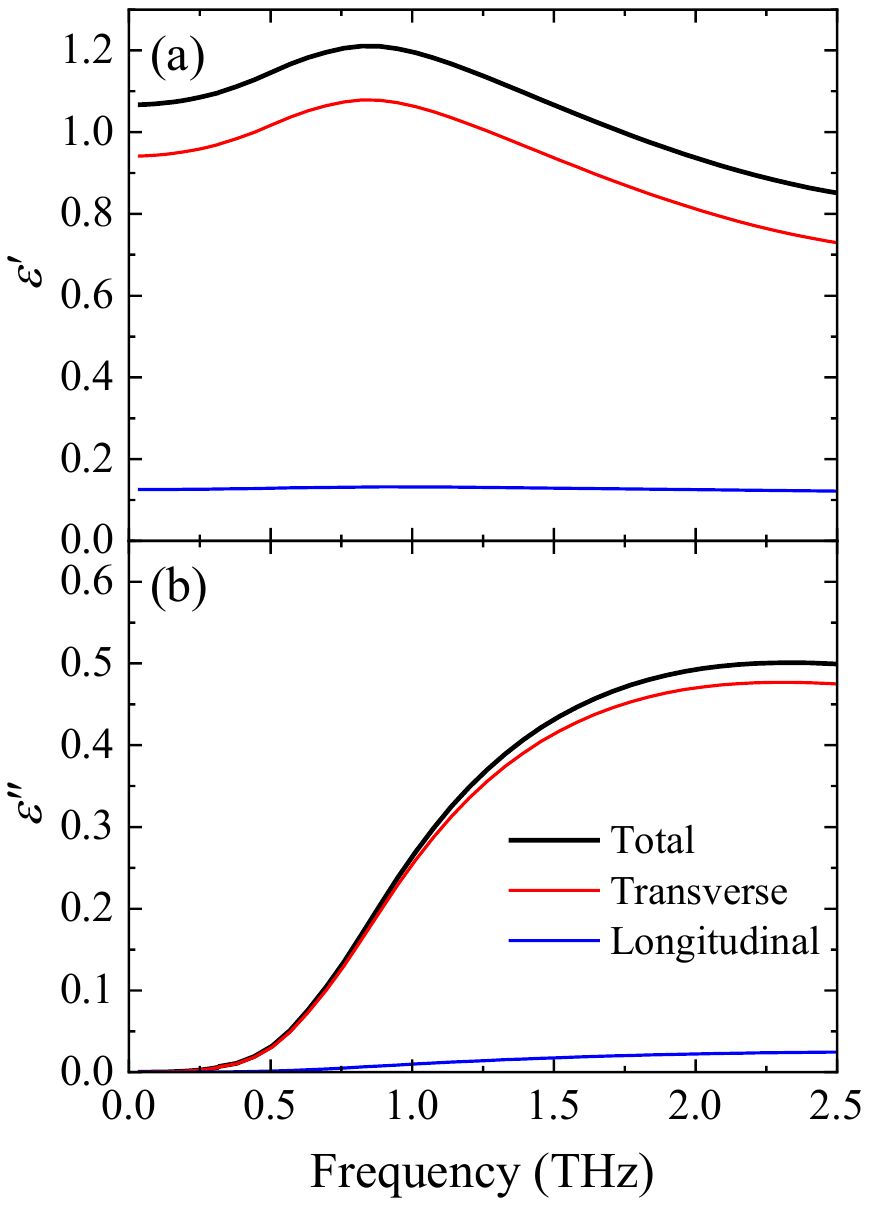} later)}
      }%
    }%
  }
\caption{Transverse and longitudinal contributions to the modeled complex dielectric function.
(a) Real part and (b) imaginary part.
Red: transverse contribution $\varepsilon_T(\omega)$; blue: longitudinal contribution $\varepsilon_L(\omega)$;
black: total $\varepsilon_T(\omega)+\varepsilon_L(\omega)=\varepsilon(\omega)-\varepsilon_\infty$.
The functions $\varepsilon_T(\omega)$ and $\varepsilon_L(\omega)$ are defined in Eqs.~\eqref{eq:appS35} and \eqref{eq:appS36}.}
  \label{fig:dielectric-decomposition}
\end{figure}

After the disorder and orientational averages introduced in Eqs.~\eqref{eq:main-disorder-correlation}
and \eqref{eq:main-polarization-averaged}, each acoustic branch carries the projection weight $1/3$.
Grouping the two degenerate transverse branches, we define $P_T(\omega)$ as their total contribution
to the field-parallel polarization and $P_L(\omega)$ as the contribution of the longitudinal branch:
\begin{equation}\label{eq:appS31}
P_T(\omega)=\frac{2E(\omega)}{3V}\sum_{\mathbf{k}}
\frac{\left\langle|\Delta q(\mathbf{k})|^{2}\right\rangle_{\mathrm{dis}}}
{-\rho\omega^{2}+G(\omega)k^{2}},
\end{equation}
\begin{equation}\label{eq:appS32}
P_L(\omega)=\frac{E(\omega)}{3V}\sum_{\mathbf{k}}
\frac{\left\langle|\Delta q(\mathbf{k})|^{2}\right\rangle_{\mathrm{dis}}}
{-\rho\omega^{2}+M(\omega)k^{2}}.
\end{equation}
The factor of $2$ in Eq.~\eqref{eq:appS31} accounts for the two degenerate transverse branches.
The total field-parallel polarization is therefore
\begin{equation}\label{eq:appS33}
P(\omega)=P_T(\omega)+P_L(\omega).
\end{equation}
Using $\varepsilon(\omega)=\varepsilon_\infty+P(\omega)/[\varepsilon_0E(\omega)]$ and applying
the Debye-sphere replacement in Eq.~\eqref{eq:main-debye-sphere} to Eqs.~\eqref{eq:appS31}
and \eqref{eq:appS32} reproduces Eq.~\eqref{eq:eps_reduced}.
We define the corresponding contributions to the modeled dielectric function as
$\varepsilon_T(\omega)=P_T(\omega)/[\varepsilon_0E(\omega)]$ and
$\varepsilon_L(\omega)=P_L(\omega)/[\varepsilon_0E(\omega)]$, so that
\begin{equation}\label{eq:appS34}
\varepsilon(\omega)=\varepsilon_{\infty}+\varepsilon_{T}(\omega)+\varepsilon_{L}(\omega),
\end{equation}
with
\begin{equation}\label{eq:appS35}
\varepsilon_{T}(\omega)=\frac{1}{\rho\varepsilon_{0}}\int_{0}^{k_{\mathrm{D}}}
\left[
\frac{2}{3}\frac{\langle|\Delta q(k)|^{2}\rangle_{\mathrm{dis}}}{-\omega^{2}+\tilde{G}(\omega)k^{2}}
\right]
\frac{3k^{2}}{k_{\mathrm{D}}^{3}}\,dk,
\end{equation}
\begin{equation}\label{eq:appS36}
\varepsilon_{L}(\omega)=\frac{1}{\rho\varepsilon_{0}}\int_{0}^{k_{\mathrm{D}}}
\left[
\frac{1}{3}\frac{\langle|\Delta q(k)|^{2}\rangle_{\mathrm{dis}}}{-\omega^{2}+\tilde{M}(\omega)k^{2}}
\right]
\frac{3k^{2}}{k_{\mathrm{D}}^{3}}\,dk.
\end{equation}
Here $\tilde{G}(\omega)$ and $\tilde{M}(\omega)$ are the mass-density-normalized moduli defined immediately below Eq.~\eqref{eq:eps_reduced}.
Figure~\ref{fig:dielectric-decomposition} summarizes the transverse and longitudinal dielectric contributions defined in
Eqs.~\eqref{eq:appS35} and \eqref{eq:appS36} by plotting their real and imaginary parts together with their sum,
$\varepsilon_{T}(\omega)+\varepsilon_{L}(\omega)=\varepsilon(\omega)-\varepsilon_{\infty}$.
Near the boson peak (BP) frequency $\omega_{\mathrm{BP}}/(2\pi)\simeq 1.0~\mathrm{THz}$, both
$\varepsilon'(\omega)$ and $\varepsilon''(\omega)$ are dominated by the transverse channel.
This suppression of the longitudinal contribution follows directly from the structure of the
longitudinal modulus, $M(\omega)=K+(4/3)G(\omega)$. In the present model, $K$ is a real,
frequency-independent constant; thus all frequency dependence and dissipation of $M(\omega)$
originate from $G(\omega)$, while the constant $K$ raises the longitudinal stiffness and
suppresses the longitudinal dielectric response.

\section{Effective shear modulus $G(\omega)$ used for glycerol}\label{app:shear-modulus}
Our dielectric model requires the complex shear modulus $G(\omega)$ as an input function.
Importantly, the present formulation does not rely on a specific microscopic origin of the BP: any
physically reasonable $G(\omega)$ that reproduces the BP part of the reduced vibrational density of
states (VDOS) is sufficient for the calculations discussed in the main text.

For glycerol, the effective shear modulus $G(\omega)$ used throughout the dielectric calculations was obtained by applying the heterogeneous-elasticity-theory coherent-potential-approximation (HET-CPA) procedure described in Ref.~\cite{Kyotani2025} to the experimental INS VDOS of glycerol glass at $80~\mathrm{K}$~\cite{Wuttke1995}.
The fixed mechanical inputs used to set the elastic and wavenumber scales were $\rho=1.26~\mathrm{g\,cm^{-3}}$, $v_{T}=1871~\mathrm{m\,s^{-1}}$, $v_{L}=3614~\mathrm{m\,s^{-1}}$, and $k_{\mathrm{D}}=1.897~\text{\AA}^{-1}$.
This numerical input set is listed for glycerol in Ref.~\cite{Kyotani2025}, and the sound velocities originate from Brillouin light-scattering measurements~\cite{Scarponi2004}.
The corresponding bulk modulus was $K=10.58~\mathrm{GPa}$, calculated as $K=\rho\left(v_{L}^{2}-4v_{T}^{2}/3\right)$.
The HET-CPA fit to the BP part of the $80~\mathrm{K}$ INS VDOS yielded $k_{\mathrm{e}}/k_{\mathrm{D}}=0.399$ ($k_{\mathrm{e}}=0.757~\text{\AA}^{-1}$), $\sigma=0.766$ ($\sigma^{2}=0.587$), and $G_{0}=5.43~\mathrm{GPa}$.
These parameters define the VDOS-constrained mechanical input $G(\omega)$ entering Eq.~(\ref{eq:eps_reduced}).

Figure~\ref{fig:shear-modulus-vdos}(a) shows the resulting real and imaginary parts of $G(\omega)$, plotted in the normalized
form $\mathrm{Re}[G(\omega)/G_{0}]$ and $\mathrm{Im}[G(\omega)/G_{0}]$.
Using the same Green-function-based relation between $G(\omega)$ and the VDOS as in Ref.~\cite{Kyotani2025},
we compute the reduced VDOS $g(\omega)/\omega^{2}$ shown in Fig.~\ref{fig:shear-modulus-vdos}(b).
The agreement with the INS spectrum demonstrates that the effective modulus used in the main-text
dielectric calculations is consistent with the experimentally observed BP spectrum.
Once the log-normal HET-CPA $G(\omega)$ used in the main analysis was obtained from the VDOS analysis, it and $K$ were kept fixed during the subsequent $(q_{0},q_{2})$ charge-parameter scans and the calculation of the dielectric response reported in the main text; no HET-CPA parameter was varied in those steps.

\section{Molecular dynamics glass configuration (glycerol)}\label{app:md-configuration}
\begin{figure}[t]
  \centering
    \IfFileExists{fig-md-configuration.pdf}{%
    \includegraphics[width=\linewidth]{fig-md-configuration}%
  }{%
    \fbox{%
      \parbox[c][0.55\textheight][c]{0.95\linewidth}{\centering
        {\small Placeholder for Fig.~\ref{fig:md-configuration} (upload \texttt{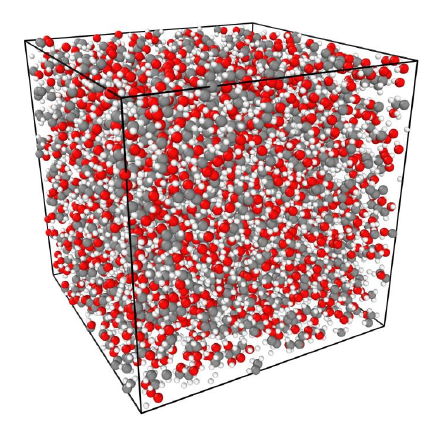} later)}
      }%
    }%
  }

  \caption{Atomic configuration of glycerol glass at $80~\mathrm{K}$ (MD).
  Representative MD snapshot of the $80~\mathrm{K}$ glycerol glass used for the structure-factor analysis
  (basis for Fig.~\ref{fig:static-structure-factors} in the main text).}
  \label{fig:md-configuration}
\end{figure}

The atomistic glycerol glass structure shown in Fig.~\ref{fig:md-configuration} was generated by classical molecular--dynamics (MD) simulation using LAMMPS. The system contains $N=12096$ atoms,
corresponding to $N_{\mathrm{mol}}=864$ glycerol molecules in an all-atom representation (14 atoms per molecule), in a cubic periodic simulation cell of side length $L=47.16~\text{\AA}$ (mass density $\rho=1.26~\mathrm{g\,cm^{-3}}$).

The interatomic interactions were described by an all-atom AMBER-based glycerol force field
following the reparameterized model of Blieck \textit{et al.}\ \cite{Blieck2005}, which is based on the AMBER glycerol
model introduced by Chelli \textit{et al.}\ \cite{Chelli1999a,Chelli1999b}. The specific parameter set employed here is
summarized in Ref.~\cite{Egorov2011}. Long-range electrostatics were treated using the particle--particle
particle--mesh (PPPM) method in LAMMPS. Starting from an equilibrated liquid configuration
at $T=350~\mathrm{K}$, the system was cooled to $T=80~\mathrm{K}$ in the canonical (NVT) ensemble at fixed
volume with a cooling rate of $0.4~\mathrm{K/ps}$ to obtain a glassy state, subsequently equilibrated at
$80~\mathrm{K}$. Figure~\ref{fig:md-configuration} shows a representative snapshot from the equilibrated $80~\mathrm{K}$ glass used for
the structural analysis. The same $80~\mathrm{K}$ configurations were used as input for the calculation of
$S_{\rho\rho}(k)$ and $S_{ZZ}(k)$ described in the following subsection.

\section{Static structure factors $S_{\rho\rho}(k)$ and $S_{ZZ}(k)$}\label{app:structure-factors}
From the MD-generated glass configuration, we evaluate the mass-density static structure factor
$S_{\rho\rho}(k)$ and the (normalized) charge--charge structure factor $S_{ZZ}(k)$.
We classify atomic sites into species (types) $\alpha,\beta$, with number fractions
$c_{\alpha}\equiv N_{\alpha}/N$ and total number density $n\equiv N/V$.
From the partial pair correlation functions $g_{\alpha\beta}(r)$ we define the total correlation functions:
\begin{equation}\label{eq:appS37}
h_{\alpha\beta}(r)\equiv g_{\alpha\beta}(r)-1.
\end{equation}
Assuming isotropy, we use the spherically averaged Fourier transform of $h_{\alpha\beta}(r)$.
In practice the real-space integral is truncated at a maximum distance $r_{\max}$ (typically
$r_{\max}=L/2$ for a cubic box of side length $L$), and we apply the standard Lorch modification
function $f_{L}(r)$ to reduce termination artifacts \cite{Lorch1969}:
\begin{equation}\label{eq:appS38}
\begin{aligned}
h_{\alpha\beta}(k)
&=4\pi\int_{0}^{r_{\max}} dr\, r^{2} h_{\alpha\beta}(r)\,\frac{\sin(kr)}{kr}\,f_{L}(r),\\
f_{L}(r)
&=\frac{\sin(\pi r/r_{\max})}{\pi r/r_{\max}}.
\end{aligned}
\end{equation}
The mass-density static structure factor is then evaluated in the Faber--Ziman form \cite{Faber-Ziman} as
\begin{equation}\label{eq:appS39}
\begin{aligned}
S_{\rho\rho}(k)
&=1+\frac{n}{\bar{m}^{2}}\sum_{\alpha,\beta} c_{\alpha}c_{\beta}m_{\alpha}m_{\beta}h_{\alpha\beta}(k),\\
\bar{m}
&\equiv \sum_{\gamma} c_{\gamma} m_{\gamma},
\end{aligned}
\end{equation}
where $m_{\alpha}$ is the atomic mass of species $\alpha$ (note that $n\bar{m}=\rho$).

For the charge--charge correlations, we use the same partial functions but weight them by the
(dimensionless) partial charges $z_{\alpha}$ (i.e., the atomic partial charge is $e z_{\alpha}$).
We follow the conventional multicomponent definition of the charge--charge structure factor
\cite[Chap.~10]{Hansen2006} and use the normalized form:
\begin{equation}\label{eq:appS40}
\begin{aligned}
S_{ZZ}(k)
&=1+\frac{n}{\bar{z}^{2}}\sum_{\alpha,\beta} c_{\alpha}c_{\beta}z_{\alpha}z_{\beta}h_{\alpha\beta}(k),\\
\bar{z}^{2}
&\equiv \sum_{\gamma} c_{\gamma} z_{\gamma}^{2}.
\end{aligned}
\end{equation}
Charge neutrality implies $\sum_{\alpha} c_{\alpha} z_{\alpha}=0$; the above normalization corresponds to
dividing the conventional $S_{ZZ}(k)$ by $\bar{z}^{2}$ and ensures $S_{ZZ}(k)\to 1$ at large $k$.

Because the pair-correlation integrals are evaluated in a finite simulation cell and truncated at $r_{\max}=L/2$, the smallest-$k$ values are sensitive to finite-size and termination effects.
We therefore exclude the points below $k=0.532~\text{\AA}^{-1}$ from the power-law analysis and use the lowest numerically reliable interval, $0.532\leq k\leq0.931~\text{\AA}^{-1}$.
A free log--log fit $S_{ZZ}(k)=A k^{n}$ over this interval gives $n\simeq4.0$, with coefficient of determination $R_{\mathrm{fit}}^{2}=0.998$. The red line in the inset of Fig.~\ref{fig:static-structure-factors} shows this fit, extended beyond the fitting interval as a guide to the eye.

\section{Parameter scan for $q_2$}\label{app:parameter-scan}
\begin{figure}[t]
  \centering
    \IfFileExists{fig-q2-scan.pdf}{%
    \includegraphics[width=\linewidth]{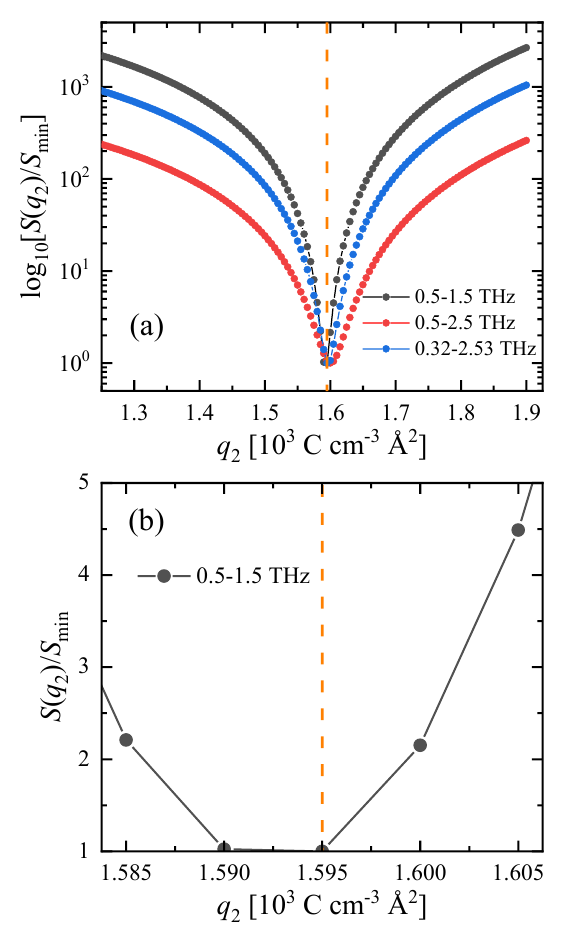}%
  }{%
    \fbox{%
      \parbox[c][0.48\textheight][c]{0.95\linewidth}{\centering
        {\small Placeholder for Fig.~\ref{fig:q2-scan} (upload \texttt{fig-q2-scan.pdf} later)}
      }%
    }%
  }

  \caption{Panel (a) shows $\log_{10}[S(q_{2})/S_{\min}]$ for three fitting windows within the experimental overlap,
  demonstrating that the minimum is robust. Panel (b) shows a zoom near the minimum for the main fitting window
  ($0.5$--$1.5~\mathrm{THz}$). The dashed line marks the adopted value $q_{2}=1.595\times 10^{3}~
  \mathrm{C\,cm^{-3}}\,\text{\AA}^{2}$.}
  \label{fig:q2-scan}
\end{figure}

In the main text, the IR-effective charge fluctuation spectrum is represented by the minimal analytic parameterization $\Delta q(k)=q_0+q_2 k^2$ over the $k$ range entering the dielectric integral, $0<k<k_{\mathrm{D}}$.
For the glycerol calculations reported here, we adopt the minimal form $\Delta q(k)=q_{2}k^{2}$ and determine $q_{2}$ by a one-parameter scan.
The role of the constant term $q_{0}$ is assessed separately by the two-parameter check
described below. For each trial value of $q_{2}$, we compute the complex permittivity
$\varepsilon(\omega)=\varepsilon'(\omega)+\mathrm{i}\varepsilon''(\omega)$ with all other model inputs fixed, including the
HET-CPA-derived mechanical input $G(\omega)$ and the values listed in
Table~\ref{tab:params}. We then evaluate a scalar misfit (sum of squared residuals)
\begin{equation}\label{eq:appS41}
\begin{aligned}
S(q_{2})=\sum_{i=1}^{N}\Bigl[
&\bigl(\varepsilon'_{\mathrm{calc}}(\omega_{i};q_{2})
-\varepsilon'_{\mathrm{exp}}(\omega_{i})\bigr)^{2} \\
&+\bigl(\varepsilon''_{\mathrm{calc}}(\omega_{i};q_{2})
-\varepsilon''_{\mathrm{exp}}(\omega_{i})\bigr)^{2}
\Bigr],
\end{aligned}
\end{equation}
where $\{\omega_{i}\}$ are the experimental frequencies within the fitting window. We focus on the BP region
(around $\sim 1~\mathrm{THz}$ for glycerol) where the vibrational contribution is prominent. The scan was
performed over $q_{2}=(1.25\text{--}1.90)\times 10^{3}~\mathrm{C\,cm^{-3}}\,\text{\AA}^{2}$ with a step of $0.005\times 10^{3}~\mathrm{C\,cm^{-3}}\,\text{\AA}^{2}$,
and Fig.~\ref{fig:q2-scan} shows the normalized misfit $S(q_{2})/S_{\min}$. A clear minimum is found at
$q_{2}=1.595\times 10^{3}~\mathrm{C\,cm^{-3}}\,\text{\AA}^{2}$ which is reported in Table~\ref{tab:params} as
$q_{2}=1.6\times 10^{3}~\mathrm{C\,cm^{-3}}\,\text{\AA}^{2}$ (rounded). Repeating the scan for slightly shifted fitting windows
within the experimental overlap changes the optimum by no more than one scan step (Fig.~\ref{fig:q2-scan}(a)).

As a check on the minimal choice $q_{0}=0$, we also minimized the same scalar
misfit with respect to both $q_{0}$ and $q_{2}$, while keeping all other model
inputs fixed, including the HET-CPA-derived mechanical input $G(\omega)$.
For the main fitting window 0.5--1.5~THz, this two-parameter minimization gives
$q_{0}=31\pm39\,\mathrm{C\,cm^{-3}}$ and
$q_{2}=(1.578\pm0.019)\times 10^{3}\,
\mathrm{C\,cm^{-3}}\,\text{\AA}^{2}$.
The uncertainties are formal one-standard-error estimates for the least-squares misfit defined above.
Thus $q_{0}=0$ lies within the fitting uncertainty.
When $q_{0}$ is fixed to zero and $q_{2}$ is re-optimized, the misfit increases by only 0.58\% relative to the two-parameter minimum. The value used in Table~\ref{tab:params}, $q_{2}=1.6\times 10^3\ \mathrm{C\,cm^{-3}}\,\text{\AA}^{2}$, is therefore
consistent with the two-parameter minimization.
We consequently retain $q_{0}=0$ as the minimal parameterization for glycerol glass at 80~K in the present frequency window.


\section{Plane-wave expansion and a microscopic route to the static $S_{ZZ}(k)$}\label{app:microscopic-szz}
\label{sec:S10_planewave_SZZ}

This section provides a microscopic interpretation of the $k$-dependent oscillator-strength spectrum
used in the continuum expression for $\varepsilon(\omega)$ in the main text.
We start from the standard linear-response mode-sum representation for the vibrational dielectric loss
in the harmonic approximation, expressed in terms of mode effective charges.
We then project disordered eigenmodes onto longitudinal and transverse plane-wave components with
mode-dependent $k$-space weights.
After displaying the four-point configurational average that appears before approximation, we introduce
a diagonal, factorized approximation in which cross/interference terms are neglected and the remaining
charge correlations are represented by the static charge--charge structure factor $S_{ZZ}(k)$.
This construction gives a microscopic route and diagnostic by which static charge correlations can
contribute to the effective coupling spectrum $\langle |\Delta q(k)|^2\rangle_{\rm dis}$ appearing in
Eq.~\eqref{eq:eps_reduced}; it is not a full microscopic determination of that spectrum over the entire
experimental frequency window.

\paragraph*{(i) Dielectric loss expressed in vibrational eigenmodes.}
We recall a standard mode-sum expression for the dielectric loss.
For an isotropic system, it can be expressed as~\cite{Thorpe1986,Pasquarello1997,Pedesseau2015}:
\begin{equation}
\varepsilon''(\omega)=\frac{4\pi^2}{3V}\sum_p\sum_{j=x,y,z}
\frac{|F^{(p)}_{j}|^2}{2\omega_p}\,\delta(\omega-\omega_p),
\label{eq:S10_eps2_modesum}
\end{equation}
where $p$ labels eigenmodes of frequency $\omega_p$ and $F^{(p)}_{j}$ is the Cartesian component of the mode charge-coupling vector (whose squared magnitude gives the oscillator strength).
In the most general form (Born effective charges) one has
\cite{Pasquarello1997,Pedesseau2015}
\begin{equation}
F^{(p)}_{j}\equiv\sum_{i,k}\frac{Z_{i,jk}}{\sqrt{m_i}}\,e^{(p)}_{i,k},
\label{eq:S10_F_Born}
\end{equation}
with $Z_{i,jk}$ the Born effective charge tensor and $\mathbf e^{(p)}_i$ the non-mass-weighted polarization vector.
For the present analysis, we adopt the rigid-ion approximation $Z_{i,jk}\simeq q_i\delta_{jk}$,
which reduces Eq.~(\ref{eq:S10_F_Born}) to the mode charge-coupling vector
\begin{equation}
\mathbf F^{(p)}\equiv\sum_i\frac{q_i}{\sqrt{m_i}}\,\mathbf e^{(p)}_i,
\qquad
\sum_{j}|F^{(p)}_{j}|^2 = |\mathbf F^{(p)}|^2.
\label{eq:S10_F_rigid}
\end{equation}
Equation~(\ref{eq:S10_eps2_modesum}) then becomes
\begin{equation}
\varepsilon''(\omega)=\frac{4\pi^2}{3V}\sum_p
\frac{|\mathbf F^{(p)}|^2}{2\omega_p}\,\delta(\omega-\omega_p).
\label{eq:S10_eps2_modesum_rigid}
\end{equation}

\paragraph*{(ii) Plane-wave expansion with explicit longitudinal/transverse polarizations.}
We project the displacement pattern of each eigenmode onto plane-wave components.
The coefficients $a_{\mathbf{k}\alpha}(\omega_p)$ specify the corresponding
$\mathbf{k}$-space weight distribution.
Taraskin \emph{et al.} explicitly state that the low-frequency
disordered modes can be approximately expanded in plane waves and that the distribution of the
plane-wave coefficients shows two relatively narrow peaks associated with transverse/longitudinal acoustic (TA/LA) hybridization
(while broadening at higher frequencies as disorder-induced attenuation becomes stronger)
\cite{Taraskin2000,Taraskin2006}.
To make the polarization content explicit, we write
\begin{equation}
\mathbf e^{(p)}_i \simeq
\sum_{\mathbf k}\sum_{\alpha\in\{L,T_1,T_2\}}
\sqrt{\frac{m_i}{\bar m}}\,
a_{\mathbf k\alpha}(\omega_p)\,
\hat{\mathbf e}_{\alpha}(\mathbf k)\,
e^{\mathrm{i}\mathbf k\cdot\mathbf r_i},
\label{eq:S10_planewave}
\end{equation}
where $\bar m\equiv N^{-1}\sum_i m_i$.
For each $\mathbf k$, the three polarization vectors form an orthonormal triad:
$\hat{\mathbf e}_{L}(\mathbf k)\parallel \mathbf k$ and
$\hat{\mathbf e}_{T_a}(\mathbf k)\cdot\mathbf k=0$ ($a=1,2$).

\paragraph*{(iii) From plane-wave expansion to a static $S_{ZZ}(k)$.}
Substituting Eq.~(\ref{eq:S10_planewave}) into Eq.~(\ref{eq:S10_F_rigid}) gives
\begin{equation}
\begin{aligned}
\mathbf F^{(p)} &\simeq
\frac{1}{\sqrt{\bar m}}
\sum_{\mathbf k,\alpha}
a_{\mathbf k\alpha}(\omega_p)\,
\hat{\mathbf e}_{\alpha}(\mathbf k)\,
\rho_Z(\mathbf k), \\
\rho_Z(\mathbf k)&\equiv \sum_i q_i e^{\mathrm{i}\mathbf k\cdot \mathbf r_i}.
\end{aligned}
\label{eq:S10_F_rho}
\end{equation}
Consider the squared projection onto the unit vector $\hat{\mathbf e}$ specifying the direction of the applied electric field.
Before the diagonal approximation is made, substituting Eq.~(\ref{eq:S10_F_rho}) and performing the
orientational average gives
\begin{align}
\left\langle |\hat{\mathbf e}\cdot\mathbf F^{(p)}|^2 \right\rangle_{\rm conf,angles}
&\simeq
\frac{1}{\bar m}
\sum_{\mathbf k,\alpha}
\sum_{\mathbf k',\alpha'}
\Gamma_{\alpha\alpha'}(\mathbf k,\mathbf k')
\nonumber\\
&\quad\times
\left\langle
 a_{\mathbf k\alpha}(\omega_p)
 a^{*}_{\mathbf k'\alpha'}(\omega_p)
\right.
\nonumber\\
&\qquad\left.
\times \rho_Z(\mathbf k)\rho_Z(\mathbf k')^{*}
\right\rangle_{\rm conf},
\label{eq:S10_projF2_fourpoint}
\end{align}
where
\begin{align}
\Gamma_{\alpha\alpha'}(\mathbf k,\mathbf k')
&\equiv
\left\langle
(\hat{\mathbf e}\cdot\hat{\mathbf e}_{\alpha}(\mathbf k))
\right.
\nonumber\\
&\qquad\left.
\times
(\hat{\mathbf e}\cdot\hat{\mathbf e}_{\alpha'}(\mathbf k'))
\right\rangle_{\rm angles}
\nonumber\\
&=
\frac{1}{3}
\hat{\mathbf e}_{\alpha}(\mathbf k)\cdot
\hat{\mathbf e}_{\alpha'}(\mathbf k') .
\label{eq:S10_Gamma_def}
\end{align}
Equation~(\ref{eq:S10_projF2_fourpoint}) shows explicitly that the general averaged oscillator strength
contains a four-point configurational average involving both the plane-wave weights and the charge-density
Fourier components.
To obtain the static-$S_{ZZ}(k)$ form used below, we make a diagonal, factorized approximation:
cross/interference terms with $(\mathbf k,\alpha)\ne(\mathbf k',\alpha')$ are neglected, and the
remaining diagonal part is factorized as
\begin{align}
&\left\langle
 a_{\mathbf k\alpha}(\omega_p)a^{*}_{\mathbf k'\alpha'}(\omega_p)
 \rho_Z(\mathbf k)\rho_Z(\mathbf k')^{*}
\right\rangle_{\rm conf}
\nonumber\\
&\quad\simeq
\delta_{\mathbf k,\mathbf k'}\delta_{\alpha,\alpha'}
\left|a_{\mathbf k\alpha}(\omega_p)\right|^2
\left\langle |\rho_Z(\mathbf k)|^2 \right\rangle_{\rm conf}.
\label{eq:S10_diag_factorized}
\end{align}
Here $|a_{\mathbf k\alpha}(\omega_p)|^2$ denotes the configuration-averaged or representative
$k$-space weight; connected correlations between the mode weights and the charge correlations are not
retained in this approximation.
Using Eq.~(\ref{eq:S10_diag_factorized}) and defining the orientational factor
$\gamma_\alpha=\Gamma_{\alpha\alpha}(\mathbf k,\mathbf k)$, we obtain
\begin{align}
\left\langle |\hat{\mathbf e}\cdot\mathbf F^{(p)}|^2 \right\rangle
&\simeq
\frac{1}{\bar m}
\sum_{\mathbf k,\alpha}
\left|a_{\mathbf k\alpha}(\omega_p)\right|^2
\left\langle |\rho_Z(\mathbf k)|^2 \right\rangle_{\rm conf}
\gamma_\alpha
\nonumber\\
&=
\frac{N\bar q^{\,2}}{\bar m}
\sum_{\mathbf k,\alpha}
\left|a_{\mathbf k\alpha}(\omega_p)\right|^2
S_{ZZ}(k)\,
\gamma_\alpha,
\label{eq:S10_projF2_SZZ}
\end{align}
where $\bar q^{\,2}\equiv N^{-1}\sum_i q_i^2$, and
\begin{equation*}
S_{ZZ}(k)\equiv \frac{1}{N\bar q^{\,2}}\left\langle \rho_Z(\mathbf k)\rho_Z(-\mathbf k)\right\rangle
\end{equation*}
is the normalized static charge--charge structure factor, consistent with the definition adopted in Appendix~\ref{app:structure-factors}
(note that $q_i=e z_i$ implies $\bar q^{\,2}=e^2\bar z^{\,2}$ in the notation used there), and
\begin{equation}
\gamma_\alpha \equiv \left\langle (\hat{\mathbf e}\cdot \hat{\mathbf e}_{\alpha})^2\right\rangle_{\rm angles}.
\label{eq:S10_gamma_def}
\end{equation}
In an isotropic three-dimensional medium the three polarizations are statistically equivalent, giving
$\gamma_L=\gamma_{T_1}=\gamma_{T_2}=1/3$ (hence a transverse-to-longitudinal weight ratio $2:1$).
If one further assumes that the two transverse polarizations are equivalent on average,
$|a_{\mathbf kT_1}|^2\simeq|a_{\mathbf kT_2}|^2\equiv |a_{\mathbf kT}|^2$, then Eq.~(\ref{eq:S10_projF2_SZZ})
can be grouped as
\begin{equation}
\begin{aligned}
\left\langle |\hat{\mathbf e}\cdot\mathbf F^{(p)}|^2 \right\rangle
&\simeq
\frac{N\bar q^{\,2}}{\bar m}\sum_{\mathbf k}
\left[
\frac{2}{3}\left|a_{\mathbf kT}(\omega_p)\right|^2 \right. \\
&\qquad\left.
+\frac{1}{3}\left|a_{\mathbf kL}(\omega_p)\right|^2
\right] \,S_{ZZ}(k).
\end{aligned}
\label{eq:S10_projF2_SZZ_LT}
\end{equation}

\paragraph*{(iv) Back to $\varepsilon''(\omega)$ and comparison with Eq.~\eqref{eq:eps_reduced} in the main text.}
Using
$\frac{1}{3}\sum_j|F^{(p)}_j|^2=\langle|\hat{\mathbf e}\cdot\mathbf F^{(p)}|^2\rangle_{\rm angles}$,
Eq.~(\ref{eq:S10_eps2_modesum_rigid}) can be viewed as a mode sum over the orientationally averaged
projection. Substituting Eq.~(\ref{eq:S10_projF2_SZZ_LT}) yields the approximate form
\begin{equation}
\begin{aligned}
\varepsilon''(\omega)&\propto
\sum_p \frac{\delta(\omega-\omega_p)}{\omega_p}
\sum_{\mathbf k}
\left[
\frac{2}{3}\left|a_{\mathbf kT}(\omega_p)\right|^2 \right. \\
&\qquad\left.
+\frac{1}{3}\left|a_{\mathbf kL}(\omega_p)\right|^2
\right] \,S_{ZZ}(k),
\end{aligned}
\label{eq:S10_eps2_SZZ_schematic}
\end{equation}
up to an overall prefactor.
Equation~(\ref{eq:S10_eps2_SZZ_schematic}) makes explicit that (i) within the diagonal,
factorized approximation of Eq.~(\ref{eq:S10_diag_factorized}), a \emph{static} $S_{ZZ}(k)$ can contribute to
the effective oscillator-strength spectrum via the plane-wave expansion, and (ii) the transverse and
longitudinal contributions appear in the $2/3$ vs $1/3$ ratio characteristic of an isotropic medium,
matching the polarization weighting used in the continuum expression of the main text [Eq.~\eqref{eq:eps_reduced}] and in
Appendices~\ref{app:disorder-average}--\ref{app:branch-decomposition}.
This static-$S_{ZZ}(k)$ construction should therefore be understood as a diagonal/factorized microscopic
route and diagnostic, not as a quantitative microscopic determination of the effective coupling spectrum
over the full $0.3$--$2.5~\mathrm{THz}$ experimental window.
This approximation is expected to be most transparent in the low-frequency propagon regime, where the $k$-space weights
have relatively narrow acoustic peaks.
Above the BP, the weights broaden as the modes become diffuson-like, and the neglected off-diagonal
terms and possible connected correlations between mode weights and charge correlations may become more
important.
A full microscopic determination of the effective charge-coupling spectrum would require a dynamic,
mode-resolved charge-correlation description, such as $S_{ZZ}(k,\omega)$ or an equivalent
eigenmode-resolved effective-charge analysis.

Finally, the physical meaning of the plane-wave weights is analogous to a $k$-space spectral function:
Taraskin \emph{et al.} report that at low frequencies the distribution of $a_{\mathbf k\alpha}(\omega)$
is dominated by relatively narrow TA/LA peaks (hybridization at fixed $\omega$), while at higher
frequencies the distribution broadens in $k$ space as disorder-induced attenuation becomes strong~\cite{Taraskin2000,Taraskin2006}.
In our continuum formalism, the corresponding broadening is captured in the imaginary part of
the effective propagators $(-\omega^2+\tilde C_\alpha(\omega)k^2)^{-1}$ appearing in Eq.~\eqref{eq:eps_reduced}.

\section{Rescaling procedure for the $R$-scan in Fig.~\ref{fig:coupling-ratio-scan}}\label{app:r-scan}

We parametrize the $k$-dependent effective charge fluctuation as
\begin{equation}\label{eq:appS52}
\Delta q(k)=q_0+q_2 k^2,
\end{equation}
and define the dimensionless ratio
\begin{equation}\label{eq:appS53}
R\equiv \frac{q_0}{q_2 k_{\mathrm{D}}^2},
\end{equation}
which measures the relative weight of the $k$-independent component at the Debye wavenumber $k_{\mathrm{D}}$.
When scanning $R$, it is useful to isolate the effect of the shape of $\Delta q(k)$ without allowing the overall fluctuation strength to grow artificially at large $R$. To this end, we rescale $(q_0,q_2)$ for each $R$ so that the Debye-sphere integrated strength is kept constant:
\begin{equation}\label{eq:appS54}
\int_{0}^{k_{\mathrm{D}}}\! dk\,k^2\,|\Delta q(k)|^2=\mathrm{const.}
\end{equation}
Using $\Delta q(k)=q_0+q_2 k^2$, the integral can be evaluated analytically as
\begin{equation}\label{eq:appS55}
\int_{0}^{k_{\mathrm{D}}}\! dk\,k^2\,(q_0+q_2 k^2)^2
= q_0^2\frac{k_{\mathrm{D}}^3}{3}+2q_0q_2\frac{k_{\mathrm{D}}^5}{5}+q_2^2\frac{k_{\mathrm{D}}^7}{7}.
\end{equation}
Substituting $q_0=R\,q_2 k_{\mathrm{D}}^2$, we obtain
\begin{equation}\label{eq:appS56}
\begin{aligned}
\int_{0}^{k_{\mathrm{D}}}\! dk\,k^2\,|\Delta q(k)|^2
&= q_2^2 k_{\mathrm{D}}^7\,J(R), \\
J(R)&=\frac{R^2}{3}+\frac{2R}{5}+\frac{1}{7}.
\end{aligned}
\end{equation}
We choose the constant to be equal to the $R=0$ value, for which $q_0=0$ and $q_2=q_2^{(0)}$.
In the calculations for Fig.~\ref{fig:coupling-ratio-scan} we used $k_{\mathrm{D}} = 1.90~\mbox{\AA}^{-1}$ and $q_2^{(0)} = 1.595\times 10^3~\mathrm{C\,cm^{-3}}\,\mbox{\AA}^{2}$ (rounded to $1.6\times 10^3$ in Table~I).
This yields the rescaling formulas
\begin{equation}\label{eq:appS57}
q_2(R)=q_2^{(0)}\sqrt{\frac{(1/7)}{J(R)}},
\qquad
q_0(R)=R\,q_2(R)\,k_{\mathrm{D}}^2.
\end{equation}
This procedure keeps the overall fluctuation strength fixed while varying the relative weight
of the $k$-independent component, thereby avoiding an artificial increase in the dielectric
strength at large $R$ and enabling a cleaner assessment of the $R$-dependence.

\section{Raman coupling coefficient of glycerol}
\label{app:raman-coupling}

Low-frequency Raman spectra of glycerol glass were measured in backscattering geometry with the confocal micro-Raman setup described in Refs.~\cite{Kabeya2016,Mori2020}. A single-longitudinal-mode, frequency-doubled diode-pumped solid-state Nd:YAG laser at $532~\mathrm{nm}$ was used for excitation; the scattered light was collected through a home-built microscope with ultranarrow-band notch filters and analyzed by an HR320 single monochromator ($1200~\mathrm{grooves/mm}$) equipped with a CCD camera (Andor DU420). Both the parallel HH and crossed HV polarization geometries were recorded, and the depolarized HV spectrum was used in the present analysis.

\begin{figure}[t]
  \centering
  \includegraphics[width=\linewidth]{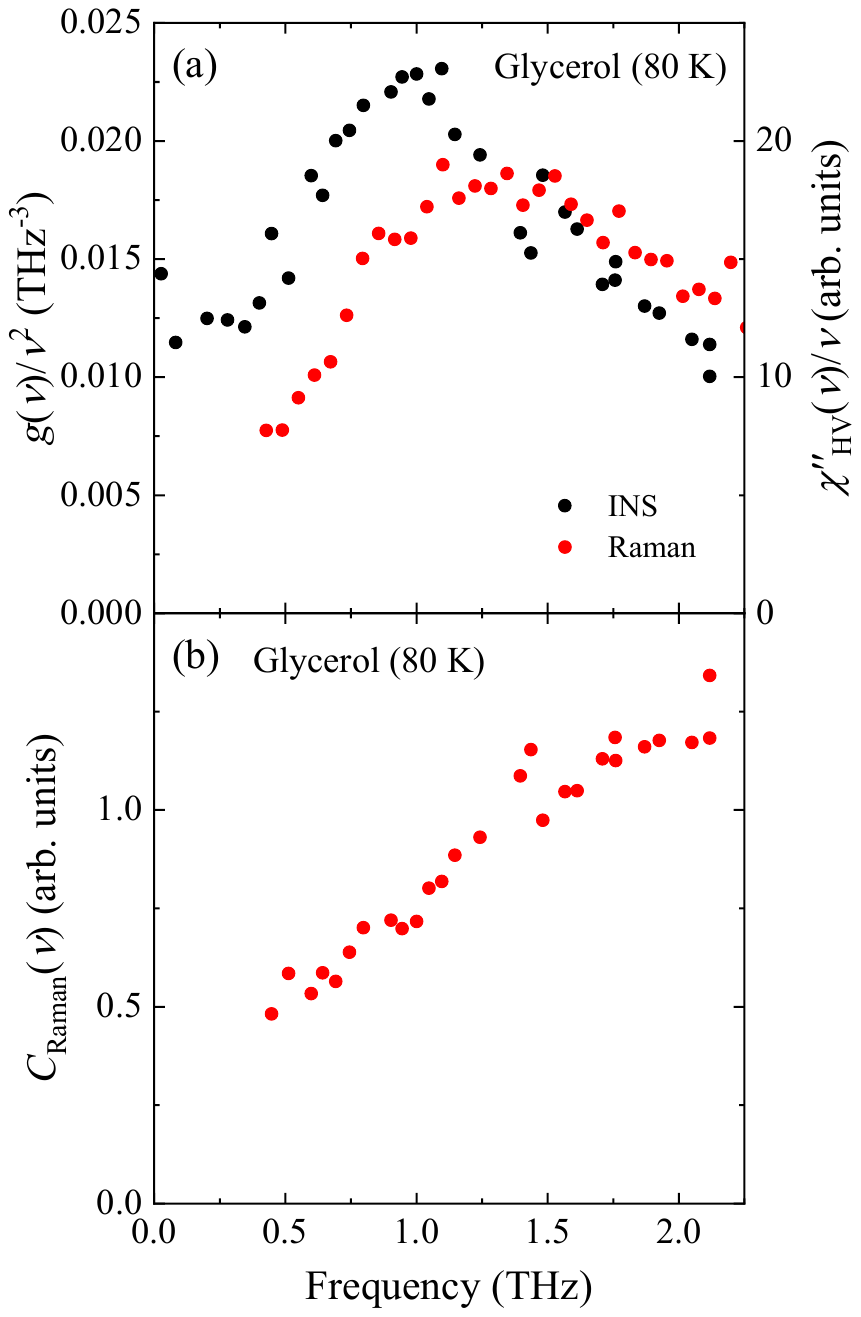}
  \caption{Raman coupling coefficient of glycerol. (a) Reduced VDOS $g(\omega)/\omega^{2}$ from the $80~\mathrm{K}$ INS data~\cite{Wuttke1995} (black symbols) and the reduced depolarized Raman response $\chi_{\rm HV}''(\omega)/\omega$ obtained from the HV spectrum measured at $80~\mathrm{K}$ (red symbols). (b) Raman coupling coefficient $C_{\rm Raman}(\omega)$ obtained by dividing the reduced Raman response by the reduced VDOS. The Raman coupling coefficient increases approximately linearly in the BP region.}
  \label{fig:raman-coupling}
\end{figure}

The Raman spectra were measured at $80~\mathrm{K}$. Following Refs.~\cite{Kabeya2016,Mori2020}, the measured Raman intensity in the HV geometry, $I_{\rm HV}(\omega)$, and the corresponding imaginary Raman susceptibility $\chi_{\rm HV}''(\omega)$ are related by $I_{\rm HV}(\omega)\propto[n_{\rm B}(\omega,T)+1]\chi_{\rm HV}''(\omega)$, where $n_{\rm B}(\omega,T)=[\exp(\hbar\omega/k_{\rm B}T)-1]^{-1}$ is the Bose--Einstein occupation number, $\hbar$ is the reduced Planck constant, $k_{\rm B}$ is the Boltzmann constant, and $T$ denotes temperature. The reduced Raman response shown in Fig.~\ref{fig:raman-coupling}(a) was therefore evaluated as $\chi_{\rm HV}''(\omega)/\omega\propto I_{\rm HV}(\omega)/\{\omega[n_{\rm B}(\omega,T)+1]\}$.

The Raman light--vibration coupling coefficient $C_{\rm Raman}(\omega)$ was obtained by combining this reduced Raman response with the $80~\mathrm{K}$ INS VDOS~\cite{Wuttke1995}, using $\chi_{\rm HV}''(\omega)/\omega=C_{\rm Raman}(\omega)g(\omega)/\omega^{2}$. Thus, over the common frequency range of the two data sets, $C_{\rm Raman}(\omega)$ is proportional to $[\chi_{\rm HV}''(\omega)/\omega]/[g(\omega)/\omega^{2}]$. As shown in Fig.~\ref{fig:raman-coupling}(b), the Raman coupling coefficient increases approximately linearly in the BP region.

\clearpage

\makeatletter\immediate\write\@auxout{\string\citation{apsrev42Control}}\makeatother
\bibliographystyle{apsrev4-2}
\bibliography{ref}

\end{document}